\documentclass[a4paper]{article}
\usepackage{blindtext, cite, array}
\usepackage{mystyle}
\usepackage[final,color]{showkeys} %%% should be loaded after hyperref        %%% add final to avoid printing the labels
\usepackage{soul}
\usepackage{dsfont}
\usepackage{graphicx}
\usepackage{subcaption}

\definecolor{refkey}{gray}{0.75}
\definecolor{labelkey}{RGB}{155,48,48}
\renewcommand*\showkeyslabelformat[1]{%
  \fbox{\parbox[t]{0.8\marginparwidth}{\raggedright\normalfont\scriptsize\url{#1}}}}

\allowdisplaybreaks
\allowbreak

\definecolor{mypurple}{RGB}{48,48,148}

\newcommand\doubleplus{{+\kern1.0ex+\kern0.8ex}}
\newcommand*{\doubleplusB}{\mathrel{\vcenter{\baselineskip1.0ex \lineskiplimit0pt
                     \hbox{\tiny+}\hbox{\tiny+}}}%
}
\renewcommand{\ddag}{\doubleplusB}

\title{From operator statistics to wormholes}

 \author[a]{Alexander Altland,}
 \author[a]{Dmitry Bagrets,}
  \author[b]{Pranjal Nayak,}
 \author[b]{Julian Sonner,}
  \author[b]{Manuel Vielma}
 \affiliation[a]{\it Institut f\"{u}r Theoretische Physik,\\ Z\"{u}lpicher Str. 77, 50937, K\"{o}ln, Germany\\}
  \affiliation[b]{\it D\'{e}partement de Physique Th\'{e}orique, Universit\'{e} de Gen\`{e}ve,\\24 quai Ernest-Ansermet, 1211 Gen\`{e}ve 4, Suisse}

\emailAdd{alexal [at] thp.uni-koeln.de}
\emailAdd{dmitry.bagrets [at] uni-koeln.de}
\emailAdd{pranjal.nayak [at] unige.ch}
\emailAdd{julian.sonner [at] unige.ch}
\emailAdd{manuel\_vielmablanco [at] hotmail.com}

\abstract{\nocite{Hawking:1974sw, Maldacena:2001kr, Barbon:2003aq}For a generic quantum many-body system, the {\it quantum ergodic} regime is defined as the limit in which the spectrum of the system resembles that of a random matrix theory (RMT) in the corresponding symmetry class. In this paper, we analyse the time dependence of correlation functions of operators. We study them in the ergodic limit as well as their approach to the ergodic limit, which is controlled by non-universal massive modes. An effective field theory (EFT) corresponding to the causal symmetry and its breaking describes the ergodic phase. We demonstrate that the resulting Goldstone-mode theory has a topological expansion, analogous to the one described in \cite{Altland:2020ccq} with added operator sources, whose leading non-trivial topologies give rise to the {\it universal ramp} seen in correlation functions. The ergodic behaviour of operators in our EFT is seen to result from a combination of RMT-like spectral statistics and Haar averaging over wave-functions. Furthermore, we analytically capture the {\it plateau behaviour} by taking into account the contribution of a second saddle point. Our main interest are quantum many-body systems with holographic duals and we explicitly establish the validity of the EFT description in the Sachdev-Ye-Kitaev (SYK) class of models, starting from their microscopic description. By studying the tower of massive modes above the Goldstone sector we get a detailed understanding of how the ergodic EFT phase is approached and we derive the relevant Thouless time scales. We point out that the topological expansion can be reinterpreted in terms of contributions of bulk wormholes and baby-universes.}

\begin{document}
%\notoc
\maketitle
%%%%%%%%%%%%%%%%%%%%%%%%%%%%%%%%%%%%%%%%%%%%%%%%%%%%%
%%%%%%%%%%%%%%%%%%%%%%%%%%%%%%%%%%%%%%%%%%%%%%%%%%%%%
%%%%%%%%%%%%%%%%%%%%%%%%%%%%%%%%%%%%%%%%%%%%%%%%%%%%%

\section{Introduction}
A resolution of Hawking's information paradox, \cite{Hawking:1974sw}, is a cornerstone to our understanding of quantum gravity. While the statement of the paradox in terms of the seemingly featureless radiation that is emitted by a black hole is the most popularly known one, the underlying problem of loss of unitarity manifests itself in many other forms. For a gravitational theory on anti-de Sitter (AdS) backgrounds, where conventionally the radiation does not escape to infinity, a natural set of observables are the boundary correlation functions. In the semiclassical approximation of the bulk theory, such correlation functions decay indefinitely at late times, \cite{Maldacena:2001kr}. This is paradoxical when viewed in light of the AdS/CFT correspondence, according to which a gravitational theory is dual to a quantum  theory living on its boundary in one lower dimension, since it can be argued \cite{Maldacena:2001kr,Barbon:2003aq} that the correlation functions cannot decay indefinitely in a quantum mechanical theory with a discrete spectrum.  Recent works have discovered new geometrical configurations of the gravitational path integral, \cite{Saad:2018bqo, Saad:2019lba, Saad:2019pqd, Almheiri:2019hni, Almheiri:2019yqk, Almheiri:2020cfm}, which relieve the tension with unitarity in observables such as correlation function or entanglement entropies. These new configurations are associated with Euclidean geometries connecting several boundaries, also known as wormholes. Given that on each of these boundaries there resides a copy of the dual quantum mechanical theory, these gravitational configurations represent connected correlations in the spectrum of the dual theory.\footnote{To see this the reader may associate each boundary with one copy of the partition function $Z(\beta)$ of the boundary theory. These connected correlations between several partition functions are tantamount to connected correlations in the spectrum of the theory.} Interestingly, these configurations also contribute to observables corresponding to single boundary geometries by introducing handles in these geometries. The type of correlations induced  by these wormhole more typically arise in situations where one considers not a single theory and its observables, but instead an {\it ensemble} of theories and an associated averaging procedure, a canonical example being random matrix theory (RMT) \cite{dyson1962a, dyson1962b, dyson1962c, mehtarmt, altland1997nonstandard}. This realisation has led to a shift in our understanding of holography in which one adopts the view that a semi-classical theory of gravity is dual to an ensemble of quantum theories, \cite{Stanford:2020wkf}. On the other hand, the authors of \cite{Altland:2020ccq} have proposed an alternative paradigm in which the ensemble description in the quantum mechanical side of the holographic duality is an {\it emergent phenomenon} that arises in physical unitary theories,\footnote{We often use the expression {\it physical unitary theory} to describe one in which an ensemble average over theories is not performed.} more specifically as a consequence of quantum chaotic dynamics. This proposal is closely related to the \emph{quantum ergodic limit} and late-time chaotic dynamics of quantum systems, as we now explain.\footnote{For some related discussion on emergence of ensemble averaging in theories also see \cite{Anninos:2016szt, Balasubramanian:2020lux, Verlinde:2021kgt, Verlinde:2021jwu}.}

%%%%%%%%%%%%%%%%%%%%%%%%%%%%%%%%%%%%%%%%%%%%%%%%%%%
%%%%%%%%%%%%%%%%%%%%%%%%%%%%%%%%%%%%%%%%%%%%%%%%%%

\subsection*{Quantum chaos and wormholes}

\begin{figure}[tbp]
\begin{center}
	\includegraphics[trim={6.5cm 8cm 7cm 4cm}, clip, width=0.66\linewidth]{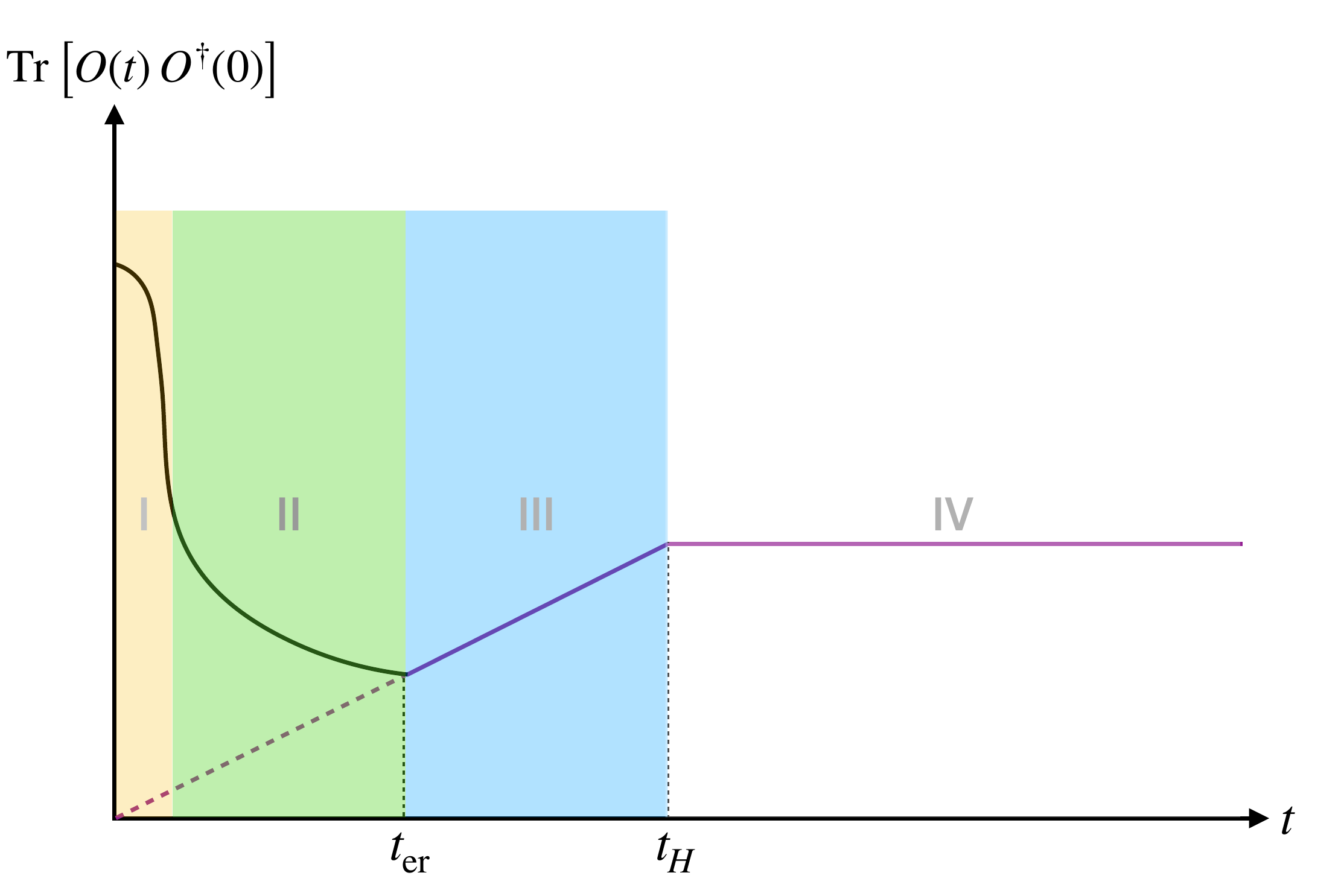}
	\caption{Late time behaviour of a generic observable in a quantum system (in the universality class of Gaussian unitary ensembles for specificity).
	From the holographic perspective, the conventional gravitational contribution computes the decay in the non-ergodic regime. Region I corresponds to the fast decay determined by the smooth energy dependence of coarse-grained DoS. This corresponds to the contribution of the disk geometry to the correlation function. In region II, correlation function decay exponentially because of presence of non-ergodic and non-universal massive modes. The part of the curve in purple is the universal ramp (region III) and plateau (region IV) behaviour and is governed by the mean level statistics of the eigenstates. The timescales at which the linear ramp becomes conspicuous is known as $t_{\rm er}$ or the ergodic time. Lastly, $t_H$ is the time at which the plateau begins and is of the order of Heisenberg time. The universal ramp-plateau corresponds to the contribution of the sum over disk geometries with handles around the standard and the Andreev-Altshuler saddle points. Also see \autoref{fig.massive-massless} for more accurate representation of operator correlation functions in the SYK model.}
\label{fig.gen-latetime}
\end{center}
\end{figure}

A key signature of quantum chaos is based on properties of the spectrum of the Hamiltonian of a quantum system. Berry and Tabor \cite{Berry-Tabor} and Bohigas-Giannoni-Schmit \cite{BGS} conjectured a remarkably universal property of quantum chaotic systems: their spectral statistics coincides with that of an appropriate RMT, while that of integrable quantum systems obeys Poisson statistics. A distinguishing feature between these behaviours is the level repulsion between the neighbouring eigenstates. For systems that lie in the symmetry class of Gaussian unitary ensembles (GUE) translating this characteristic behaviour of level spacings into the time domain, in other words following the time evolution of correlation functions and similar probes, generates another universal signature: a linearly rising ramp, culminating in a late-time plateau, as further illustrated in \autoref{fig.gen-latetime}.

The onset of the linear growth, also referred to as \emph{ramp}, is known as the \emph{ergodic/Thouless time}, $t_{\rm er}/t_{\rm Th}$. This time scale marks the onset of the universal behaviour that can be attributed to the RMT-like spectral properties in a given chaotic quantum system. In a single realisation, the ramp-plateau are superimposed with fluctuations and the universal behaviour emerges as an average over these fluctuations. At the level of the spectrum of the theory, this universal behaviour manifests itself in the two-point correlation function of the density of states (DoS),
\begin{equation}\label{eq.sine-kernel}
\left\langle \rho\left(-\frac\omega2\right) \rho\left( \frac\omega2 \right) \right\rangle =
\pi \delta\left(\frac{\pi\omega}{\Delta}\right) + 1 - \frac{\sin^2\left(\pi\omega/\Delta\right) }{ \left(\pi\omega/\Delta\right)^2 } =:  
\pi \delta\left(\frac{\pi\omega}{\Delta}\right) + 1 - {\rm sinc}^2\left(\frac{\pi\omega}\Delta \right).
\end{equation}
The occurrence of this sine-kernel, ${\rm sinc}(\omega)$ gives rise to the universal ramp-plateau behaviour. Here, $\Delta$ is the mean level-spacing and is defined precisely below.

Operator expectation values in pure states or thermodynamic ensembles form a more physical set of observables in a theory as they can be measured directly in experiments. For example, two-point functions of operators are the simplest observables measuring the linear response in a system. A perpetually decaying correlation function is characteristic of a coupling to an external system and hence non-unitarity. Therefore it is important to understand how the transition in the behaviour of a correlation function from an initial decay to a non-vanishing plateau at long times happens. Additionally, operators are more appropriate observables to understand how a system in a pure state might appear to be thermal, \`{a} la Eigenstate thermalisation hypothesis (ETH), \cite{Deutsch, Srednicki, DAlessio:2016rwt, Sonner:2017hxc, Nayak:2019khe, Nayak:2019evx}. The observable, the {\it operator resolvent}, that our work focuses on in this paper is related to many such physical observables quite simply, as explained in \autoref{sec.op-res}. One of the main results of this paper is to establish a similar sine-kernel behaviour for many-body operators, $O$, governed by the operator sine kernel
\begin{equation}\label{eq.OSK}\begin{aligned}
		\left\langle { O}\left(-\frac\omega2\right) { O}^\dagger\left( \frac\omega2 \right) \right\rangle &\sim \Bigg( \pi \delta \left(\frac{\pi\omega}\Delta\right) + 1 - {\rm sinc}^2\left(\frac{\pi\omega}\Delta \right)  \Bigg) \left({\rm Tr}\left[ O O^\dagger \right]-\frac1D {\rm Tr}[ O] {\rm Tr}\left[ O^\dagger\right] \right) \\
	 &\qquad +  \delta \left(\frac{\pi\omega}\Delta\right) \ {\rm Tr}[ O] {\rm Tr}\left[ O^\dagger\right]~.
\end{aligned}\end{equation}
It is interesting that the `1' in \eqref{eq.sine-kernel} corresponds to the disconnected part of the correlation function whereas in \eqref{eq.OSK} it is part of the connected part of the correlation function. As we explain in detail in the bulk of this work, the traces on the right hand side are over the entire Hilbert space or parts of it. Fourier transforming into the time domain gives rise to the universal ramp-plateau behaviour of operator correlations illustrated in \autoref{fig.gen-latetime}. Another observation is that this result is consistent with the form of the ETH ansatz of \cite{Deutsch, Srednicki}, $O_{\alpha\beta} = \bar O(E) \, \delta_{\alpha \beta} + e^{-S(E)/2} f( E,\omega) \, R_{\alpha \beta}$, where $\bar O$ is the expectation value of the operator in a microcanonical ensemble at the energy, $E$; $S(E)$ is the corresponding entropy in the microcanonical ensemble, and $R_{\alpha \beta}$ is a random number of zero mean and unit variance.  Our result here encapsulates universal contributions to ETH-like correlations in the ergodic phase of {\it any} chaotic quantum system, as well as ergodic and non-ergodic contributions for the Sachdev-Ye-Kitaev (SYK) model in particular. The emergence of the ergodic phase in a wide class of quantum many-body systems was recently studied in \cite{Altland:2020ccq, Altland:2017eao, Monteiro:2020fpa}. The relevant degrees of freedom in the ergodic phase correspond to certain pseudo-Goldstone modes that arise from the causal symmetry breaking in the study of correlation functions of the kind \eqref{eq.sine-kernel}.  This effective field theory (EFT) description applies in the late time (ergodic) limit, while at early times  (non-ergodic limit), observables decay at a rate determined by the smooth energy dependence of the coarse-grained DoS, $\rho(E)$, of the system, and are insensitive to the finer details of the spectrum. In between these two regimes lies a non-universal intermediate regime which we analyse below for cases of interest in holographic duality.

In the bulk the initial coarse-grained decay corresponds to the prediction of semiclassical gravity. What then corresponds to the late-time universal ergodic behaviour? The na\"ive semiclassical gravity computation reproduces the initial decay of physical observables, but it does not reproduce the ramp-plateau behaviour;  the observables instead decay to zero. Remarkably, including new wormhole configurations in the semi-classical gravity path integral reproduces the required linear ramp at late times. Our EFT description of the ergodic phase of a holographic many-body theory allows us to predict these contributions and to classify them systematically. We establish that the behaviour of the system in this regime is equivalent to that of a random-matrix theory where the matrix is interpreted as the Hamiltonian of the system. One may then interpret each diagram that arises in the EFT description in a topological expansion as constructing a bulk surface of given topology, along the lines of the old matrix models \cite{Kazakov1985, KAZAKOV1988171}. For the case of SYK and similar low-dimensional examples one can literally interpret this construction as the triangulation of the bulk geometry, but for higher-dimensional examples the bulk interpretation remains somewhat more mysterious.

 At even later times, the plateau arises due to doubly non-perturbative corrections. From our discussion above, this means that such configurations in the semi-classical gravity contain information about the fine-grained spectrum of the quantum theory, albeit in an averaged sense. That is to say, they don't reproduce the noisy fluctuations that are observed in a physical unitary theory, but do reproduce their averaged behaviour.

We would like to highlight a fact that might be of particular holographic interest. Our work establishes that the wormhole geometries receive contributions not only from the spectral details of the theory as described in \eqref{eq.sine-kernel}, but also from the Haar integration of the unitaries in an ensemble description. The causal-symmetry approach advocated here thus provides a unified description of the different field-theoretic descriptions of the gravitational wormhole configurations that have been put forward in \cite{Saad:2019lba, Saad:2019pqd, Altland:2020ccq} and \cite{Pollack:2020gfa}. Another important feature that has emerged and that we explicitly corroborate in this work is the fact that the gravitational configurations that contribute to the observables in the ergodic phase (at late times) depend upon the observables under study, in contrast to what was widely believed originally.

\subsection*{Going beyond the ergodic regime}
In the pre-ergodic phase, the above-mentioned transition from an initial decay to the universal ergodic regime happens at intermediate timescales  governed by the `massive modes' of the theory. As discussed in detail in \cite{Altland:2020ccq} and in the present work in the specific context of the SYK model, these massive modes correspond to Hilbert space non-homogeneous projections of the collective field. The ramp starts when the lightest of these modes decay at $t_{\rm er}$. Reflecting the non-universal nature of this regime, the ergodic time itself may vary for different observables, as is apparent by comparing the results of this work with those of \cite{Altland:2017eao}.

\subsection*{Ergodic regime in physical theories}
Having argued for the interest and importance of spectral chaos in a quantum system as well as in a dual theory of gravity, we want to emphasise that demonstrating these behaviours in a given field theory is a non-trivial task. For example, quantum ergodicity has not been explicitly demonstrated for $\mathcal N=4$ SYM, which is a benchmark of holographic quantum field theories. In fact, the study of the spectral statistics of quantum systems has largely been a numerical endeavour. The SYK model, \cite{Sachdev:1992fk, Kitaev-talks:2015, Maldacena:2016hyu}, is a toy-model of $N$ Majorana fermions that facilitates explicit analytic computations. Hence, it provides an excellent laboratory to test and improve our understanding of many of these issues related to quantum chaos. Since it is dual to a two-dimensional theory of gravity, \cite{Maldacena:2016upp, Jensen:2016pah, Engelsoy:2016xyb, Mandal:2017thl}, it also sheds light on various aspects of AdS/CFT correspondence in general. In fact, it was the original observation by Kitaev, \cite{Kitaev-talks:2015}, namely that the OTOCs in this model grow exponentially with a Lyapunov index that saturates the bound proposed in \cite{Maldacena:2015waa} which has led to intensive study of this model in recent times. It might appear that the SYK model is already the same as a RMT since it is a theory with a Hamiltonian chosen from an ensemble, i.e. a theory of disorder averaging. However, for the SYK model with a $(q/2)$–body interaction and a Hilbert space of dimension, $D=2^{N/2}$, only ${N\choose q} \sim N^q$ matrix of elements out of a total of $D^2$ in the Hamiltonian are chosen randomly. This exponentially small amount of randomness indeed leads to the deviations from the RMT behaviour seen at early times. Quantum ergodicity in the SYK model was studied numerically in \cite{Cotler:2016fpe, Gharibyan:2018jrp}; and subsequently, analytically in \cite{Altland:2017eao, Micklitz:2019qlm, Garcia-Garcia:2021elz}. In these works, the late time behaviour of the spectral form factor is studied using two different techniques, namely the replica trick and supersymmetry, and the emergence of RMT-like statistics is derived analytically. The study of the spectral form factor as a fine-grained probe of the spectrum in mesoscopic physics is quite old, \cite{Altshuler:1986, Andreev:1995}; however, it was popularly introduced in the string theory community originally in \cite{Papadodimas:2015xma} and more recently in \cite{Cotler:2016fpe}. In the present work, we generalise their results for the case of operator two-point functions. Before we proceed further, let us provide a brief structure of this paper along with a preview of our main results.

\subsection*{Main results and the structure of the paper}
We discuss the ergodic limit of the SYK model in \autoref{sec.erglimSYK}. In the process, we study the so-called operator resolvent which measures the behaviour of an operator two-point function in energy eigenstates as a function of the difference of their energies. This is closely related to finite time thermal two-point functions as discussed in \autoref{sec.op-res}. In the rest of the section, we provide a path integral representation of this observable and derive the precise ergodic limit in the case of the SYK model. This is done by studying the model in the energy (or equivalently, time) regime when the universal $\sigma$-model physics becomes important. At late times, the average behaviour of the operator correlation functions has a ramp and the plateau behaviour that originates from the contribution of the Goldstone modes in the sigma model. Unlike the observables such as the spectral resolvent, in this case the relevant contribution comes not only from the spectral statistics but also from averaging over the SYK eigenstates, which in the ergodic limit becomes equivalent to the Haar averaging. Importantly, our result also clarifies the relationship between ETH and RMT predictions for the operator correlation functions. Not only do these observables depend on the behaviour of operators in energy eigenstates, they also contain the information about the distribution of the eigenstates themselves as a function of the energy.

In \autoref{sec.massive}, we study the slowest non-universal non-ergodic modes that mark the late time transition of observables from non-universal to universal ergodic behaviour, the timescales corresponding to the onset of the RMT behaviour. An important observation is that $t_{\rm er}$ depends on the observables under study.

We finally conclude with some discussion in \autoref{sec.summary}. For a more in-depth discussion of the supersymmetric methods we point the reader to \cite{Altland:2020ccq, efetov-book, haake-book, Verbaarschot_review}. In \autoref{app.saddle}, we discuss the dominance of the different saddle point solutions of the classical equation of motion. Various details of the derivation of the $\sigma$-model action and the computation of operator correlators in the ergodic limit that have been omitted in the main part of the paper are presented in \autoref{app.details}. In \autoref{app.massive} we discuss the details of the non-ergodic modes that govern the physics at Thouless time.

%%%%%%%%%%%%%%%%%%%%%%%%%%%%%%%%%%%%%%%%%%%%%%%%%%%%%
%%%%%%%%%%%%%%%%%%%%%%%%%%%%%%%%%%%%%%%%%%%%%%%%%%%%%
%%%%%%%%%%%%%%%%%%%%%%%%%%%%%%%%%%%%%%%%%%%%%%%%%%%%%

\section{Ergodic limit and \texorpdfstring{$\sigma$}{Sigma}-model}\label{sec.erglimSYK}
We begin this section with an introduction of the observables that we are interested in. The operator resolvents
\begin{equation}\label{eq.resolvent}
	R(\omega) = \sum_{\alpha,\beta} \left| \left\langle \alpha| O|\beta \right\rangle\right|^2 \delta\left(E_\alpha-E_\beta-\omega\right)~,
\end{equation}
measure the probability of all quantum mechanical processes that exchange a given amount of energy in the presence of the operator insertion. In other words, they quantify the behaviour of the operator matrix elements in energy eigenbasis (state $|\alpha\rangle$ is an eigenstate with energy, $E_\alpha$) corresponding to the states with a fixed energy difference across the spectrum. Below, we describe how the operator resolvents that we study in this work are related to the time dependent correlation functions in (micro-)canonical ensembles. In a general setup, we write a generating function as a path integral over some auxiliary variables to compute these observables.

\subsection{Operator resolvent}\label{sec.op-res}
The Fourier transform of the resolvent is related to the thermal two-point function of the operator $O$ at infinite temperature,
\begin{equation}\label{eq.infty-temp-corr}
	C(t) = {\rm Tr}\left[ O(t)  O^\dagger(0) \right] = \int \!\! d\omega\, e^{i \omega t} R(\omega)~.
\end{equation}
Note that the energy argument, $\omega$, is dual to the real time parameter, $t$. Therefore, early/late times correspond to large/small values of $\omega$, and are referred to as UV/IR, respectively. This is different from the other notion of UV/IR which refers to the low/high energy part of the spectrum of a theory.
From the spectral resolvent one can obtain other more familiar quantities, such as

\paragraph{The correlation function in the Canonical Ensemble:} this is the Fourier transform of the spectral resolvent of an operator of the form $O \to e^{-\frac\beta2 H} O$ at temperature, $T = 1/\beta$.

\paragraph{The correlation function in the Microcanonical Ensemble:} this quantity can be obtained by an insertion of a projection operator, $\mathcal P_{\mathcal W}$, that restricts the sum over the eigenstates to an appropriate microcanonical window, $E_{\alpha, \beta} \in \left(\bar E - \Delta E, \bar E+\Delta E \right) =: \mathcal W$. This can be achieved by replacing $O\to O \mathcal P_{\mathcal W}$, $O^\dagger\to O^\dagger \mathcal P_{\mathcal W}$. Note that the new insertions aren't Hermitian conjugates of each other. A possible choice of $\mathcal P_{\mathcal W} = e^{-\frac\beta2H}$ gives the regulated thermal two-point function,
\begin{equation}\label{eq.regulated-thermal-corr}
	{\rm Tr}\left[ e^{-\frac\beta2H} O(t) e^{-\frac\beta2H} O(0) \right]~.
\end{equation}
This regulated correlation function is related to the canonical correlation function by an analytic continuation, $t\to t + i \frac\beta2$, and is relevant for the observables in the holographic setting. Note that any operator can be written as a sum over a traceless and a trace component,
\[
	O = {\mathcal O} + \frac {{\rm Tr}[O]} D \mathds 1, \quad {\rm Tr}[\mathcal {O}] = 0~.
\]
The analysis of the term proportional to identity needs to be performed more carefully. Subsequently, for pedagogical reasons we show the details of only the traceless term. However, the full answer including the traceful and the traceless parts is presented in \eqref{eq.final-sigma.tracefull}.

\noindent The operator resolvent can be written in an integral representation that is more useful for our purpose,
\begin{align}\label{eq:resolvent}
	R(\omega)  &=\frac 1{2\pi^2} \int dE \;{\rm Re}\!\left[ {\rm Tr}_{\mathcal H} \left[ G^+\!\!\left(E+ \tfrac{\omega}{2}      \right) { \mathcal O}~ G^-\!\!\left(E- \tfrac{\omega}{2}\right) { \mathcal O}^\dagger \right] \right. \nonumber \\
			  & \qquad \qquad \qquad - \left.  {\rm Tr}_{\mathcal H} \left[ G^+\!\!\left(E+ \tfrac{\omega}{2}\right) { \mathcal O}~ G^+\!\!\left(E- \tfrac{\omega}{2}\right) { \mathcal O}^\dagger \right]\right] \\
			  &=: \frac1{2\pi^2} \int dE \,\Big( R^\pm(E,\omega) - R^{\ddag} (E,\omega)\Big), \nonumber
\end{align}
where $G^\pm(E) = (E \pm i 0^+ - H)^{-1}$ are the resolvents of the many-body Hamiltonian.\footnote{Recall, that the density of states is related to the Green's functions by, $\rho(E) = \mp{\rm Im Tr}G^\pm(E)$, and the above combination reproduces the product of density of states.}
Henceforth, we concentrate on these two terms separately. The second term above doesn't contain the ergodic modes but needs to be kept in order to get the correct factors to obtain level-repulsion and is studied in Appendix \ref{app.plusplus}. The non-trivial physics of the $\sigma$-model arises from the first term. It can be rewritten as a path integral,
\begin{align}\label{eq:resolvent.2}
	&\int dE \;{\rm Re}\!\left[ {\rm Tr}_{\mathcal H} \left[ G^+\!\!\left(E+ \tfrac{\omega}{2}\right) { \mathcal O}~ G^-\!\!\left(E- \tfrac{\omega}{2}\right) { \mathcal O}^\dagger \right] \right] \\
	&= \int dE \;{\rm Re}\!\left[ \partial_{h_+}\partial_{h_-} Z[ h]\right] \Big|_{h_\pm=0} =: \int \!\!dE\,R^\pm(E,\omega) \nonumber~,
\end{align}
where $Z[ h]$ is the generating function for the correlation functions. One can use either the replica trick \cite{Altland:2017eao} or supersymmetric techniques \cite{Micklitz:2019qlm} to compute the generating function. A recent work by some of the authors, \cite{Altland:2020ccq}, discusses these techniques quite generally. We refer to \autoref{app.details} for the details of the computations presented in this section. $Z[ h]$ can be written as a graded Gaussian integral,
\begin{equation}\label{eq.Gen-func}
	Z[ h] =  \int \mathcal D \bar \Psi \mathcal D \Psi \exp\left[ i\bar \Psi \cdot \left(  z -  H +  h \right) \cdot \Psi \right]~.
\end{equation}
Here, $\bar \Psi, \Psi$ are $4D$ dimensional {\it auxiliary} graded vectors with $2D$ Grassmann components and $2D$ c-number components acting in the tensor product ${\rm R/A} \, \otimes\, {\rm b/f}\, \otimes {\cal V}$ of three linear spaces corresponding to the retarded/advance, boson/fermion and the physical many-body Fock space of the SYK model, respectively. The retarded/advanced product space corresponds to the choice of the sign of the regulator in the kernel of the above integral, (2.6). This in turn ensures that the Gaussian integration over corresponding components of the graded vector, $\Psi$, gives a differently regulated Green’s function, $G^\pm(E) = (E \pm i 0^+ - H)^{-1}$. This is needed for generating both the factors of $G^\pm$ in (2.4) using the integration of $\Psi$ fields. Also, $\bar \Psi = \Psi^\dagger \cdot \sigma_3^{\rm RA}$, which guarantees a convergence of the Gaussian path integral~(\ref{eq.Gen-func}) in the advanced sector. The variable $ z$ is defined as,
\begin{equation*}
	 z = E  + \left(\frac\omega2 + i 0^+ \right) \sigma_3^{\rm RA}, 
\end{equation*}
where the Pauli matrix $\sigma_3^{\rm RA}$ acts in the ${\rm RA}$-space as indicated by the upper index.
The kernel for the Gaussian integral contains the Hamiltonian of the theory under study, $ H$, and appropriate sources,
\begin{equation}\begin{aligned}\label{eq.src}
	 h =  \begin{bmatrix}
		0& h_+\, P_b\otimes \mathcal O\\
		-h_- \,P_b\otimes \mathcal O^\dagger&0
	\end{bmatrix}_{\rm RA}~.
\end{aligned}\end{equation}
Here $P_b$ is the projector to the bosonic sector, while a block structure refers to the ${\rm RA}$-space. It can be readily checked that \eqref{eq:resolvent.2} holds.
The determinant factors arising from the Gaussian integral cancel between the Grassmann and the c-number sector once the sources have been switched-off, $h_\pm=0$, resulting in a normalised correlation function.

\begin{figure}
\begin{subfigure}{.5\textwidth}
  \centering
  \includegraphics[width=.8\linewidth]{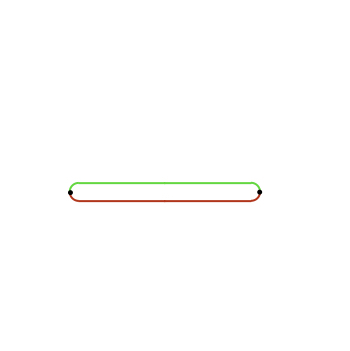}
%  \caption{This Feynman diagram corresponds to the leading contribution to the operator 2-point correlation function.}
  \caption{}
  \label{fig:sfig1}
\end{subfigure}%
\begin{subfigure}{.5\textwidth}
  \centering
  \includegraphics[width=.8\linewidth]{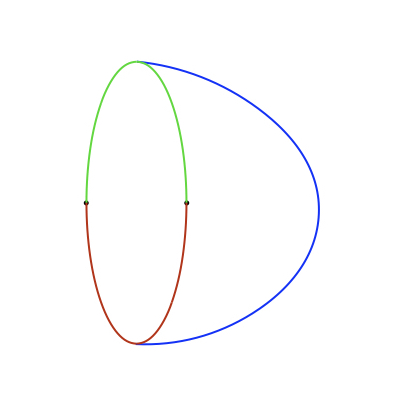}
%  \caption{The depiction of the Feynman diagram in (a) as a bulk geometry, which is a disk in this case.}
  \caption{}
  \label{fig:sfig2}
\end{subfigure}
%\caption{Diagrams corresponding to the leading order contribution to the operator 2-point function. Here, the differently coloured lines demonstrate different causalities. The dots represent the projectors on the bosonic sector as well as the insertion of the sources for the operators. It arises from computing the resolvent with $\ddag$-causality in \eqref{eq:resolvent} and therefore does not receive contributions from the ergodic modes, $B,\tilde B$. This diagram can be understood to arise from the integration over the eigenfunctions of the Hamiltonian. In this diagram we have refrained from explicitly demonstrating the disorder average. The bulk is understood to {\it emerge} out of disorder averaging.}
  \caption{The leading contribution to the operator 2-point correlation function arises from the diagram demonstrated in (a). Here, the differently coloured lines demonstrate different causalities. The dots represent the projectors on the bosonic sector as well as the insertion of the sources for the operators. It arises from computing the resolvent with $\ddag$-causality in \eqref{eq:resolvent} and therefore does not receive contributions from the ergodic modes, $B,\tilde B$. This diagram can be understood to arise from the integration over the eigenfunctions of the Hamiltonian. In this diagram we have refrained from explicitly demonstrating the disorder average. The figure (b) is a depiction of the diagram as a bulk geometry, which is a disk in this case. The bulk is understood to {\it emerge} out of disorder averaging.}
\label{fig.leading}
\end{figure}

\begin{figure}
\begin{subfigure}{.5\textwidth}
  \centering
  \includegraphics[width=.8\linewidth]{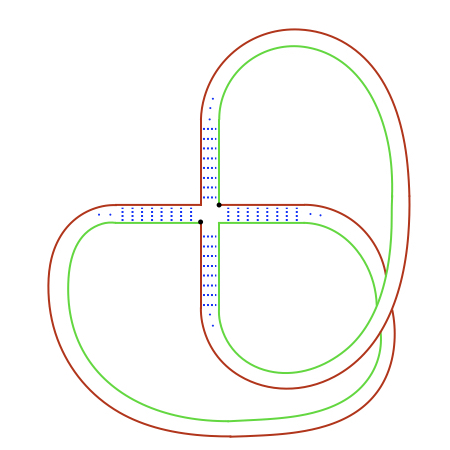}
%  \caption{This Feynman diagram corresponds to the leading contribution to the operator 2-point correlation function.}
  \caption{}
  \label{fig:sfig3a}
\end{subfigure}%
\begin{subfigure}{.5\textwidth}
  \centering
  \includegraphics[width=.8\linewidth]{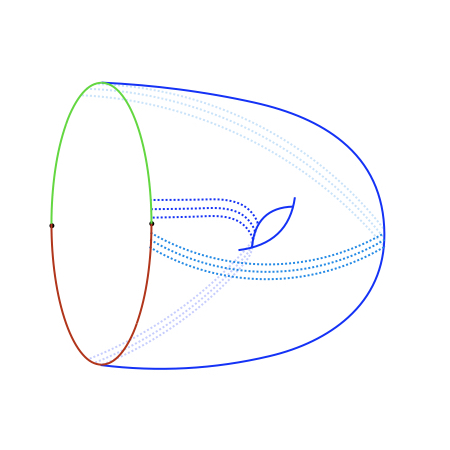}
%  \caption{The depiction of the Feynman diagram in (a) as a bulk geometry, which is a disk in this case.}
  \caption{}
  \label{fig:sfig3b}
\end{subfigure}
%\caption{Diagrams corresponding to the leading order contribution to the operator 2-point function. Here, the differently coloured lines demonstrate different causalities. The dots represent the projectors on the bosonic sector as well as the insertion of the sources for the operators. It arises from computing the resolvent with $\ddag$-causality in \eqref{eq:resolvent} and therefore does not receive contributions from the ergodic modes, $B,\tilde B$. This diagram can be understood to arise from the integration over the eigenfunctions of the Hamiltonian. In this diagram we have refrained from explicitly demonstrating the disorder average. The bulk is understood to {\it emerge} out of disorder averaging.}
  \caption{The subleading contribution to the operator correlation function arises from the non-planar diagram of figure (a). The coloured fat-line propagators demonstrate matrix fields $B,\tilde B$, respectively, in the double line notation (refer to \eqref{eq.pertT-param} below). These $B,\tilde B$ fields have one index each in the advanced and the retarded sector, which is demonstrated by different colours of the double lines. The dots represent the projectors on the bosonic sector as well as the insertion of the sources for the operators. The blue dotted lines are the Hamiltonian exchange, or the disorder average. We invite the reader to refer to \autoref{app.details} for a microscopic understanding of the above diagrams. This diagram is dual to a bulk geometry demonstrated in (b): a torus with a boundary. The dotted blue lines span the two dimensional surface, {\it constructing} the bulk geometry, albeit this time they wind around the different cycles of the genus-1 Riemann surface.}
\label{fig.subleading}
\end{figure}

As described here, the above technique is quite general, although, it is a little formal and abstract. It provides little insight into how to obtain the spectral properties of a physical unitary system. An integral over the elements of the Hamiltonian in \eqref{eq.Gen-func} provides an analytical handle to compute an average over such physical observables described above, \cite{efetov-book, haake-book, Sedrakyan:2020oip, Verbaarschot_review}. The RMT was originally developed to model the strongly interacting Hamiltonians that arise in nuclear and atomic systems. The important question is to understand how it arises as an effective description in a physical unitary theory. The high frequency quantum noise that exists in physical unitary systems over and above the smooth averaged behaviour easily dephases under any form of averaging, not just that of the Hamiltonian. An averaging over even small energy scales leads to dephasing. This is reminiscent of the RG interpretation where coarse-graining over small timescales (equivalently, lengthscales) scales or large energy scales leads to an effective theory. In \cite{Altland:2020ccq} it was argued that the emergence of this effective description is captured by a low-energy $\sigma$-model. Therefore, instead of searching for the randomness of the Hamiltonian in physical systems, one should try to understand in what physical regime the $\sigma$-model describes our strongly-interacting systems like $\mathcal N=4$ SYM theory. Demonstrating it for chaotic holographic field theories like SYM is still a daunting task, however there exist holographic many-body systems such as the SYK model which do provide a viable proving ground to demonstrate these ideas. As explained in the introduction, the SYK model, despite being defined as an ensemble of random Hamiltonians, is not a RMT. Its Hamiltonian has much smaller randomly chosen independent elements than the full RMT. We now specialise our calculations to the SYK model as a demonstration of our ideas. Some general lessons that we learn from the sigma model approach for the operator correlations functions in the ergodic limit of any theory can be easily summarised as follows. The leading contribution is due to planar diagrams with a single boundary as depicted in \autoref{fig.leading}.\footnote{The terms \emph{leading} and \emph{subleading} are used to refer to the first and the second terms in the operator sine kernel, \eqref{eq.OSK}. While this nomenclature is true for a typical light operator it might not be true for a general operator.} We have drawn this diagram in the \emph{flavour space}. Alternatively it can be drawn in the \emph{colour space} where the blue lines that demonstrate the disorder average themselves are drawn in a double line notation corresponding to the colour-indices carried by the Hamiltonian. In this dual description the colour lines can be directly interpreted as triangulating a two-dimensional surface with a disk topology (see \cite{Altland:2020ccq} for a more detailed discussion on the colour-flavour duality). This is reminiscent of the seminal work on the duality between a particular double-scaled matrix model and 2-dimensional gravity, \cite{Kazakov1985, KAZAKOV1988171}.\footnote{See the reviews \cite{Ginsparg:1993is, DiFrancesco:1993cyw, Klebanov:1991qa} for a comprehensive list of references, and the reference \cite{Gaiotto:2003yb} for an example in minimal string theory.} The difference between the present case and the models studied in the past is that the matrix models in the present case are only an effective description captured by the Gaussian matrix models in the late-time limit. Recent work of \cite{Goel:2021wim} might provide a more concrete connection in the case of the SYK model but we also believe that this observation holds for an arbitrary quantum theory which might not have an explicit matrix model description.

The subleading contribution to the operator correlation function is given by a torus with a boundary, depicted in \autoref{fig.subleading}. This diagram contributes the sine kernel, \eqref{eq.OSK}, which gives rise to the ramp-plateau behaviour. It is important to note that in terms of the geometries the Euclidean wormholes directly contribute to the variance of the physical observables as discussed in \cite{Saad:2018bqo, Saad:2019lba}. More generally, the number of boundaries of the bulk geometry are related to the number of traces in the physical observables. However, wormholes still contribute as intermediate states (baby universes) in other observables. Within the sigma model this is manifested by the properties of the Wick contraction, \eqref{eq.BBtprop}. The first of these equations can be seen as a contribution of the wormhole, while the second one can be seen as an emission of baby universes.

\subsection{SYK model}
The SYK model, \cite{Sachdev:1992fk, Kitaev-talks:2015, Maldacena:2016hyu}, is a (0+1)-dimensional theory of $N$ Majorana fermions with the Hamiltonian,
\begin{equation}\label{eq.SYKHamiltonian}
	 H = \sum_{1\le i_1<\ldots<i_4\le N}\!\!\!\!J_{i_1i_2i_3 i_4} \psi^{i_1}\ldots \psi^{i_4} \equiv \frac1{4!} \sum_{i_1,\ldots,i_4=1}^N\!\!\!\!J_{i_1i_2 i_3 i_4} \psi^{i_1}\ldots \psi^{i_4}~,
\end{equation}
The coupling constants, $J_{i_1i_2 i_3 i_4}$, are chosen randomly from a Gaussian ensemble,
\begin{equation}
	P\left(J_{i_1\cdots i_4} \right) \sim \exp\left[-\frac12 \frac{N^{3}}{3! \,J^2} J^2_{i_1\cdots i_4}\right]~.
\end{equation}
Note that the SYK Hamiltonian is quite sparse, there are only $n= {N\choose 4}$ \emph{independent} matrix elements in it out of the total $D^2 = 2^{N-2}$ elements of the full Hamiltonian. This, in principle, is the reason behind the deviations from random matrix theory. For brevity, we denote the product of four Majorana fermions by
\begin{equation}\label{eq.basis-sparse}
	X^a \equiv \psi^{i_1} \psi^{i_2} \psi^{i_3} \psi^{i_4},
\end{equation}
where, $a \equiv \{i_1,i_2,i_3, i_4\}$ is a condensed index labelling the $q$-fermions in the individual interaction terms of the Hamiltonian and satisfy the condition $1\le i_1 < i_2 < i_3 < i_4 \le N$. Consequently, $\left\{ X^a \right\}$ is the basis of matrices that span the space of sparse SYK Hamiltonians. We also denote a basis of arbitrary $D\times D$ matrices by a product of arbitrary number of Majorana fermions,
\begin{equation}\label{eq.basis-full}
	X^{\mu_k} \equiv \psi^{i_1}\ldots \psi^{i_k}~.
\end{equation}
The index $\mu_k \equiv \{i_1, i_2, \ldots, i_k \}$ not only labels the individual fermions that enter the product but also the number of fermions ($k$) in the product itself. It therefore satisfies the constraints $1\le i_1 < i_2 < \ldots < i_k \le N$ for $k = 0, 1, \ldots, N$. The element $X^0 = \mathds 1$ corresponding to $k=0$. In case of theories with an ensemble average, \eqref{eq.Gen-func} is modified to include the disorder average,
\begin{equation}\label{eq.Gen-func-dis}
Z[ h] = \left\langle \int \mathcal D \bar \Psi \mathcal D \Psi \exp\left[ i\bar \Psi \cdot (z - H  + h) \cdot\Psi \right] \right\rangle_{\rm dis}~,
\end{equation}
where, $\langle \cdot \rangle_{\rm dis}$ denotes the disorder average.
On substituting the SYK Hamiltonian in \eqref{eq.Gen-func-dis} and performing the disorder average, one generates a quartic interaction term in the graded super-vector $\bar \Psi, \Psi$. This quartic term can be reduced to a quadratic term by introducing an auxiliary field, $A^a$, using Hubbard-Stratonovic transformation. The remaining path integral is \footnote{we are using the notation $X^\mu$ for both the matrix in the $D$-dimensional Hilbert space, as well as $\mathds1^{\rm RA}\otimes\mathds1^{\rm bf}\otimes X^\mu$. The exact usage is clear from the context. Also, $A^a$ are $4D$-dimensional matrices.}
\begin{equation}\label{eq.Z-disordered-avg-Hubb-Strat}
	Z[ h] \! =\!\! \int\!\! \mathcal D \bar \Psi \mathcal D \Psi \mathcal D A \exp\bigg[  i\bar\Psi\cdot  \left(  z +  h \right)\cdot \Psi  -\frac1{2n} \sum_a {\rm STr}\left[X^a A^a X^a A^a\right] + \frac{i\gamma} n \sum_a{\rm STr}\left[\Psi\bar\Psi A^a\right]\bigg]~.
\end{equation}
We have introduced the energy scale
\begin{equation}
	\gamma = \frac12 J\sqrt N~,
\end{equation}
in the large $N$ limit. This is a quadratic path integral in the vectors, $\bar \Psi, \Psi$, which can be integrated out to obtain,
\begin{equation}
\label{eq.Z_h_As}
	Z[h] \! =\!\! \int\!\! \mathcal D A \exp\bigg[  -\frac1{2n} \sum_a {\rm STr}\left[X^a A^a X^a A^a\right] - {\rm STr} \ln \bigl[   z +  h + \frac{\gamma}{n} \sum_a A^a \bigr]\biggl]
\end{equation} 
We decompose the Hubbard-Stratanovic fields into average and difference fields,
\begin{equation}\label{eq.com}
	A^a = \bar A + \mathcal A^a, \qquad \bar A = \frac 1 n \sum_a A^a, \qquad \sum_a \mathcal A^a = 0~.
\end{equation}
Since only $\bar A$ enters the logarithmic part of the action in~(\ref{eq.Z_h_As}), the path integral over difference fields $\mathcal A^a$ is Gaussian and can be completed exactly.\footnote{The Jacobian of the integration over the difference fields is trivial because they are related to the original fields only through linear transformations. Additionally because the constraint on the sum of these fields in \eqref{eq.com} is also linear, it doesn't give rise to any Jacobian. The one-loop determinant is trivial because these fields are graded fields and the integration is quadratic (this follows from Parisi–Sourlas–Efetov–Wegner theorem, \cite{efetov-book, haake-book}). For the details of these integrals we refer the reader to Appendix A of \cite{Altland:2017eao}.} The field $\bar A$ can be written using the basis \eqref{eq.basis-full},
\begin{equation}\label{eq:decomposition}
	\bar A = \sum_\mu \bar a^\mu X^\mu~.
\end{equation}
Here, $\bar a^\mu$ carries the 4 dimensional indices in the supersymmetry and adv./ret. space, while $X^\mu$ carries the Hilbert space indices.\footnote{We have simplified the notation so that $\mu$ collectively refers to the number of Majorana fermions in the basis, $k$, as well as all possible distinct choices of them for a fixed number.} In terms of these fields, the generating function is
\begin{equation}\label{eq.fullaction}
	\int \!\! \mathcal D \bar a^\mu \; \exp\left[ -\frac D2 \sum_{\mu} s(\mu)S(\mu)^{-1} \; {\rm STr} \left[ \bar a^\mu \cdot \bar a^\mu \right] - {\rm STr}[\ln(K)]\right]~,
\end{equation}
where, $K$ denotes the kernel of the quadratic integral over the superfields, $\Psi$ and $\bar \Psi$,
\begin{equation*}
	K = \Big(  z +  h + \gamma \bar A \Big)~.
\end{equation*}
Note that the ${\rm STr}$ represents super-trace over the 4-dimensional space in the first term, but a super-trace over $4D$-dimensional space in the second term. In future, we refrain from explicitly stating this fact as it will always be clear from the context. The permutation symbols $s(\mu), S(\mu)$, count the sign when permuting $X^\mu$ across themselves or $X^a$. Specifically, $s(\mu) = \pm 1$ is defined by $(X^\mu)^\dagger = s(\mu) X^\mu$ and depends only on the number of Majoranas $|\mu|$ contained in $\mu$. The sum $S(\mu)$ reads
\begin{equation}
	S(\mu) = \frac 1 n \sum_a s(\mu,a), \qquad n={N\choose 4},
\end{equation}
where sign factors $s(\mu,a) = \pm 1$ stem from commutation relations $X^\mu X^a = s(\mu,a) X^a X^\mu$. At this stage, we make an ansatz that will be justified retrospectively in \autoref{sec.massive}. We assume
\begin{equation}\label{eq.hha}
	\bar a ^\mu =  y \, \delta^{\mu,0}~,
\end{equation}
i.e. the dominant saddle point solution(s) are homogeneous in the Hilbert space. We refer to this ansatz as \emph{homogeneous Hilbert space ansatz} (HHA). With this ansatz, the path integral in \eqref{eq.fullaction} is reduced to an integral only over the homogenous modes,
\begin{equation}\label{eq.HHA-action}
	Z_0[h]=\int \! \mathcal D  y \; \exp\left[ -\frac D2  {\rm STr}(y^2) - {\rm STr} \ln(K_0) \right], \qquad
	K_0 = \Big(  z +  h  + \gamma \;  y \otimes \mathds{1}^{\rm Fock} \Big)~,
\end{equation}
where $\mathds{1}^{\rm Fock}$ stands for the identity operator in the physical Fock space.

\subsection{Causal symmetry breaking}
The path integral in \eqref{eq.HHA-action} can be studied in the classical saddle point approximation when $D\to\infty$. The dominant solution of the saddle-point equations is
\begin{equation}\label{eq.saddle-sol}
	 y_0 = - \frac E{2\gamma} + i \Lambda \sqrt{ 1 - \frac {E^2}{4\gamma^2}}, \qquad
	\Lambda = \sigma_3^{\rm RA} \otimes \mathds{1}^{\rm bf}.
\end{equation}
A more detailed study of choosing the correct saddle point out of the 16 na\"{i}ve possibilities can be found in any standard text on the subject like \cite{efetov-book, haake-book}. The procedure is also succinctly reviewed in \autoref{app.saddle}.
 
Note that $ y \to T\cdot  y \cdot T^{-1}$ is the symmetry of the first term in \eqref{eq.HHA-action} for all $T \in U(1,1|2)$. However, it is a symmetry of the full action only if $T\cdot  z \cdot T^{-1} =  z$. For $\omega = 0, \varepsilon\to0$, the full $T \in U(1,1|2)$ remains the symmetry of the action which is weakly (and explicitly) broken by $\omega \neq 0$. The saddle point solution, \eqref{eq.saddle-sol}, breaks this symmetry spontaneously to $U(1|1)\times U(1|1)$. The coset space $U(1,1|2)/U(1|1)\times U(1|1)$ is that of pseudo-Goldstones which have a finite action cost associated with them when $\omega\neq0$. However, this action cost is suppressed with respect to the saddle point action by a factor of $\omega/\gamma\sim \omega/(J\sqrt{N})$.Thus the perturbative treatment is valid when
\begin{equation}
	\frac \omega{\gamma} \ll 1 \Rightarrow \frac\omega{\Delta} \ll D~,
\end{equation}
where $\Delta \sim\gamma/D$ is the average many-body level spacing in the band center. This provides an explicit realisation of the universal causal symmetry breaking that was shown in \cite{Altland:2020ccq} to give rise to the ergodic behaviour more generally.

\noindent The sigma-model manifold generated by the $T$-transformations contains an alternate saddle point of the equations of motion corresponding to, \cite{Andreev:1995},
\begin{equation}\label{eq.aa-saddle}
	\Lambda_{AA} = \sigma_3^{\rm RA} \otimes \sigma_3^{\rm bf}~,
\end{equation}
which can be arrived at from the dominant saddle point by a special transformation,\\ $T_0\in U(1,1|2)/U(1|1)\times U(1|1)$,
\begin{equation}\label{eq.aa-saddle_transform}
	T_0 = \mathds{1}^{\rm RA} \otimes P_b +  \sigma_1^{\rm RA}  \otimes P_f, \qquad T_0 \cdot \Lambda \cdot T_0^{-1} = \Lambda_{AA},
\end{equation}
where $P_b$ and $P_f$ denote the projectors to the boson/fermion sectors, respectively. 
We refer to this saddle point as the \emph{Andreev-Altshuler saddle point} subsequently. The contribution of this saddle point is non-perturbative with respect to the contribution of the dominant saddle, \eqref{eq.saddle-sol}. Ideally, one should integrate over the entire sigma-model manifold because these modes are `soft'. However, because of the one-loop exactness of the theory, it is sufficient to sum the perturbative contributions around the individual saddle points, \cite{efetov-book, haake-book}.

\subsection{Integral on the \texorpdfstring{$\sigma$}{sigma}-model manifold}
Following the above discussion, when the integration over the $y$-variable is reduced to the $\sigma$-model manifold, the generating function is given by the following path integral,
\begin{equation}\begin{aligned}
\label{eq.path-pert-ds}
Z[h] =	\int \mathcal D Q &\exp\left[ \frac {i \pi \rho(E)}{2}\,  {\rm Str}\left( \omega \Lambda Q\right) - S_{\rm src}(h) \right], \qquad Q = T \Lambda T^{-1}.
\end{aligned}\end{equation}
$Q$ is a point on the coset space $U(1,1|2)/U(1|1)\times U(1|1)$. We refer the reader to \autoref{app.details} for the details of the computation on how to derive this equation from \eqref{eq.HHA-action}. Here, $\rho(E)$ is the mean-field density of states given by,
\begin{equation}
\label{eq.Wigner_DoS}
\rho(E) = \frac {D}{\pi \gamma} \sqrt{1-\frac{E^2}{4\gamma^2}}\,.
\end{equation}
This is a good approximation of the density of states only near the centre of the band of the SYK spectrum. For modifications in the present analysis to give the correct DoS away from the band-centre see \cite{Altland:2021toappear}. The action $S_{\rm src}(h)$ contains all terms up to quadratic order in the sources, which are 
obtained by expanding the log-term in \eqref{eq.HHA-action} w.r.t. $h_\pm$., see details in Appendix~\ref{app.details}.
On differentiating the generating function~(\ref{eq.path-pert-ds}) twice, one gets the building block of the correlation function~(\ref{eq:resolvent.2}),
\begin{equation}\label{eq.path.integral.interest}
\partial_{h_+}\partial_{h_-} Z[ h]\Bigl|_{h_\pm=0} = \int \mathcal D Q \, \exp\left[ \frac {i \pi \rho(E)}{2}\,  {\rm Str}\left( \omega \Lambda Q\right)\right] \left(\partial_{h_+} S_{\rm src}\times \, \partial_{h_-} S_{\rm src}- \partial_{h_+}\partial_{h_-}S_{\rm src}\right)\Bigl|_{h_\pm=0}~.
\end{equation}
The path integral above features the semi-classical exactness and thus can be evaluated by performing the Gaussian integration around the two saddle points, $\Lambda$ and $\Lambda_{AA}$.
In the first case the $T$ matrix within the perturbation theory can be parametrised as
\begin{equation}\begin{aligned}\label{eq.pertT}
	T &= \exp(-W) \approx \mathds{1} - W + \frac1{2!}W^2 + \cdots \\
\end{aligned}\end{equation}
where we introduced the matrices
\begin{equation}\label{eq.pertT-param}
W = - \begin{pmatrix}0&B\\\tilde B&0\end{pmatrix}, \qquad B=\begin{pmatrix}
		x&\mu\\\nu& iy
	\end{pmatrix}, \qquad \tilde B = \begin{pmatrix}
		x^*&\bar\nu\\
		-\bar\mu& iy^*
	\end{pmatrix}\,.
\end{equation}
Here $x,y \in \mathds{C}$, while $\mu, \bar\mu, \nu$ and $\bar\nu$ are independent Grassmann variables.
In analogy with low-energy QCD, one may think of $W$ as the pions as described for example in \cite{Altland:2020ccq}.
For the contribution around the Andreev-Altshuler saddle point $\Lambda_{AA}$ one parametrise $T = T_0 \exp(-W)$,
where $W$ is defined as before.

\subsubsection{Perturbations around the dominant saddle point}
It can be verified that it is sufficient to expand the action in \eqref{eq.path.integral.interest} to quadratic order in $W$ and the sources in the pre-exponent to quartic order. The action then becomes
\begin{equation}
S[Q]  = - \frac {i\pi \rho(E)}{2}\,  {\rm Str}\left( \omega \Lambda Q\right) =  - 2 i  \pi \rho(E) \omega\, {\rm Str} \left[ B\tilde B \right] + \dots
\end{equation}
It can be regarded as the ``quark mass'' for the pion field $W$.
An astute reader might point out that it is not consistent to expand to different orders. However, it can be checked that the contribution arising from the quartic interaction term in the exponent evaluates to zero. This is related to the non-renormalisation theorem for the GUE symmetry class, \cite{efetov-book}, the only one we treat explicitly in this paper.
The expression for sources expanded to higher order in $W$ can be found in \autoref{app.details}. Performing the integrals on the pion fields and taking the real part to compute the resolvent, we obtain
\begin{equation}
	R^\pm_0(E,\omega) = - \frac{\pi^2\rho^2(E)}{s^2} \,\frac{{\rm Tr}\left[\mathcal O \mathcal O^\dagger \right]}{D^2}
	 + \frac{{\rm Tr}\left[\mathcal O \mathcal O^\dagger \right]}{\gamma^2}.
\end{equation}
We have used the dimensionless frequency,
\begin{equation}
	s = \pi \rho(E)\omega = \frac{\pi \omega}{\Delta(E)}.
\end{equation}
This concludes our evaluation of the contribution from the standard saddle point to the operator resolvent. To this we need to add the contribution from the Andreev-Altshuler saddle point, which we shall do now.

\subsubsection{Perturbations around the Andreev-Altshuler saddle point}
The Gaussian action around the Andreev-Altshuler saddle point~\eqref{eq.aa-saddle} evaluates to
\begin{equation}
S_{AA}[Q]  =   -2 i \pi \rho(E)\omega  - i \pi \rho(E)\omega \, {\rm STr}\left[ \sigma_3^{\rm bf} B\tilde B + \sigma_3^{\rm bf} \tilde B B \right].
\end{equation}
Note that the subleading saddle point $\Lambda_{AA}$ corresponds to the Weyl symmetry, which is equivalent to exchanging the  two eigenvalues in the fermionic sector of the standard saddle point $\Lambda$.We refer the reader to \cite{Altland:2020ccq} for a detailed discussion on this symmetry in terms of exchanging the north and the south poles of the $S^2$ factor of the coset manifold. The evaluation of the resolvent gives
\begin{equation}
	R^\pm_{AA}(E,\omega) = \pi^2 \rho^2(E) \Bigg( 2 \pi  \delta(s) + \frac {\cos(2s)} {s^2} \Bigg)
	\frac{{\rm Tr}\left[\mathcal O \mathcal O^\dagger \right]}{D^2}.
\end{equation}
Adding together the contributions of both the saddle points, the total resolvent is
\begin{equation}\begin{aligned}
	R^\pm(E,\omega) &=  2 \pi^2  \rho^2(E) \left( \pi \delta(s) - \frac {\sin^2(s)} {s^2} \right)\frac{{\rm Tr}\left[\mathcal O \mathcal O^\dagger \right]}{D^2}
	+ \frac{{\rm Tr}\left[\mathcal O \mathcal O^\dagger \right]}{\gamma^2}.
\end{aligned}\end{equation}

\begin{figure}
\begin{subfigure}{.5\textwidth}
  \centering
  \includegraphics[width=.95\linewidth]{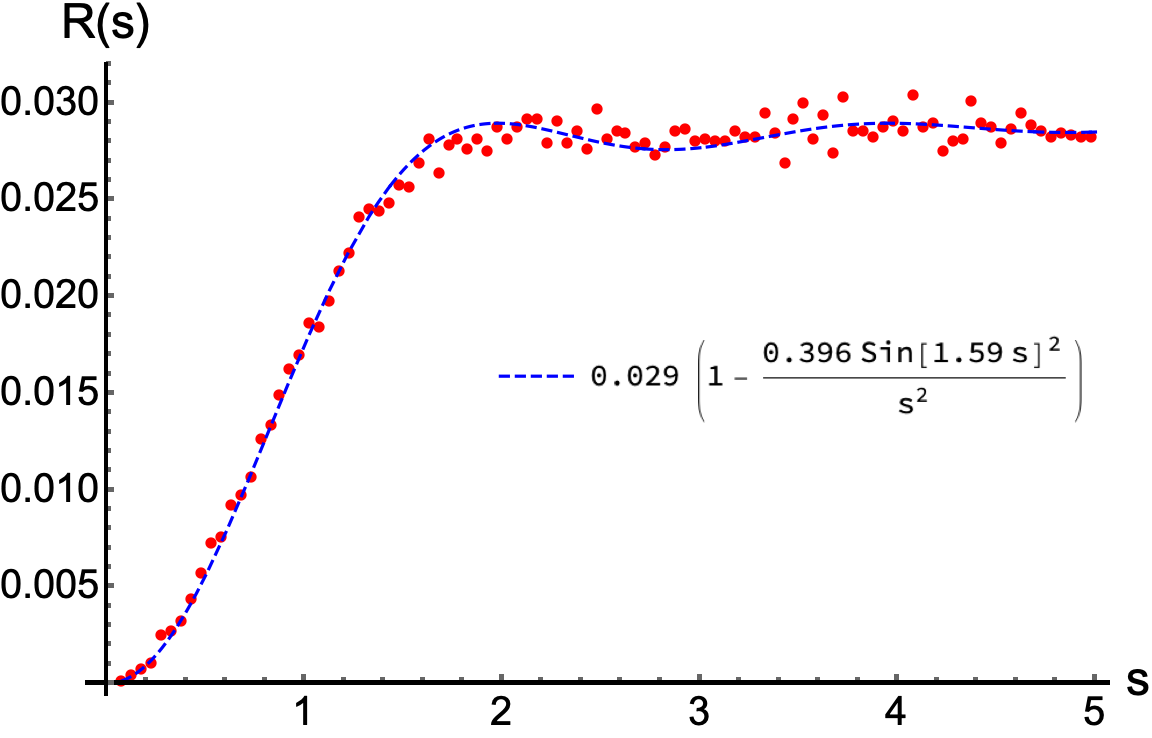}
  \caption{$N = 14$}
  \label{fig:sfig4a}
\end{subfigure}%
\begin{subfigure}{.5\textwidth}
  \centering
  \includegraphics[width=.95\linewidth]{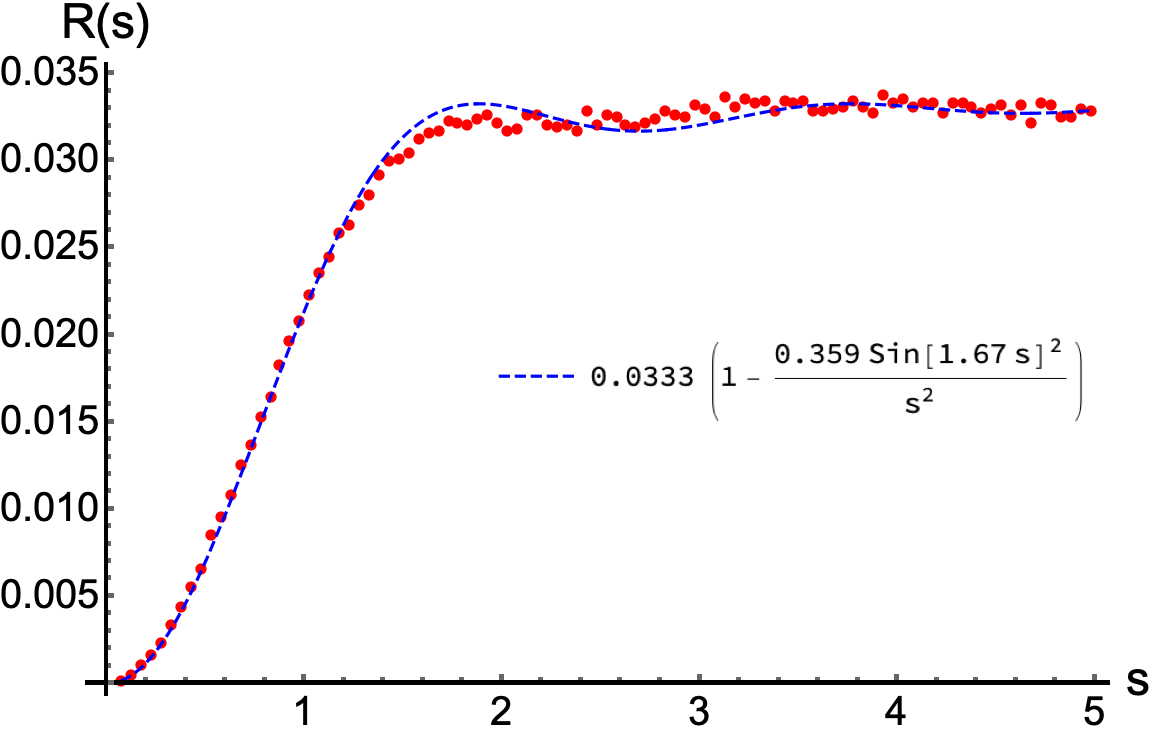}
  \caption{$N = 18$}
  \label{fig:sfig4b}
\end{subfigure}
  \caption{Comparing the numerically computed value of the resolvent as defined in \eqref{eq.resolvent} with the prediction of the $\sigma$-model for the SYK model with $N$ Majorana fermions. In (a), for $N = 14$ we have computed the resolvent for \emph{hopping operator} between \emph{sites} 3 and 5. In (b), for $N=18$, the resolvent of the \emph{hopping operator} between \emph{sites} 4 and 6 is computed. In our conventions, a system of $N$ Majorana fermions has $N/2$ \emph{sites}.}
\label{fig.numerics}
\end{figure}

\noindent To get the final answer we need to subtract from the above the contribution of $R^{\ddag}$ computed in \eqref{eq.res-plusplus},
\begin{equation}\label{eq.final-sigma}
	R(E,\omega) = R^\pm(E,\omega) - R^{\ddag}(E,\omega) = 2  \pi^2 \frac{\rho^2(E)}{D^2}  \left( \pi \delta(s) +1- \frac {\sin^2(s)} {s^2} \right) {\rm Tr}\left[\mathcal O \mathcal O^\dagger \right].
\end{equation}
However, recall that so far we have been working with traceless operators. Generalising to the case of traceful operators requires a more careful analysis and we present only the final result here,
\begin{align}\label{eq.final-sigma.tracefull}
	R(E,\omega) &=2  \pi^2 \frac{\rho^2(E)}{D^2} \left[ \pi \delta (s) \ \bigl|{\rm Tr}\,O \bigr|^2 
	+ \left( \pi \delta(s) +1- \frac {\sin^2(s)} {s^2} \right)\left( {\rm Tr}\left[{O} {O}^\dagger \right] - \frac 1 D \bigl|{\rm Tr}\, O \bigr|^2 
	\right)\right]~.
\end{align}
This is the operator resolvent in the ergodic limit, \eqref{eq.OSK}, advertised in the introduction. \autoref{fig.numerics} compares the numerical computation of the operator resolvent in the SYK model with this analytical porediction. Before we consider the contributions of the non-ergodic modes in the next section, let us look at the behaviour of the correlators as a function of time.

\subsection{Late time behaviour}
As discussed in the introduction, the important feature of the observables in energy space that leads to the universal ramp-plateau behaviour at late times is the appearance of the sine-kernel. However, to see this behaviour explicitly one needs to perform additional integrals of the resolvent, \eqref{eq.final-sigma.tracefull}, over the energies $E,\omega$,
\begin{align}
	C_{\rm RMT}(t) &= \frac1{2\pi^2}\int \!\!dE \,d\omega \,e^{i\omega t} R(E,\omega) =
	\frac1{2\pi^3} \int\!\!dE\,ds\, \Delta(E) \,e^{is \cdot t \frac{\Delta(E)}{\pi}} \,R(E,s) \nonumber \\
	& = \frac{1}{4 D^2}\int \!\!dE\ \rho(E) \bigg[ 4  {\rm Tr}\, O \, {\rm Tr} \, O^\dagger 
	+ 
	\left( {\rm Tr}\left[{O} {O}^\dagger \right] - \frac 1 D \bigl|{\rm Tr}\, O\bigr|^2 
	\right) \nonumber \\
	&\times
	\left( 2 + t \, \frac{\Delta(E)}{\pi} + \left(2- t\,\frac{\Delta(E)}{\pi} \right)\, \text{sgn}\left(t\, \frac{\Delta(E)}{\pi} -2\right) \right)\bigg]~.
\end{align}

The time at which the ramp transitions into the plateau depends on the value of energy, $E$, of the microcanonical window because $\Delta(E)$ is a function of $E$. Therefore, the above integral is a convolution of various ramps and plateaus defined for each microcanonical energy window. The last ramp ends at the Heisenberg time $t_H = 2D/\gamma$ corresponding to the band center. 
When working with the infinite-temperature canonical ensemble, the integration with the DoS~(\ref{eq.Wigner_DoS}) leads to the answer
\begin{equation}\label{eq.CRMT}\begin{aligned}
	C_{\rm RMT}(t) &=\left\{ \begin{matrix}
		 \displaystyle\frac1D \bigl|{\rm Tr} \,O \bigr|^2
		+ \frac 2 {\pi D} \left(\tilde t \sqrt{1-\tilde t^2}+\arcsin \tilde t  \right) 
		\left({\rm Tr}\left[{O} {O}^\dagger\right]  - \frac1D \bigl|{\rm Tr} \,O \bigr|^2\right), &\quad \tilde t < 1 \\[15pt]
		\displaystyle
		 \left(\frac1D - \frac1{D^2}\right) \bigl|{\rm Tr} \,O \bigr|^2 + \frac1D {\rm Tr}\left[{O} {O}^\dagger\right] , &\quad \tilde t > 1
	\end{matrix}\right.
\end{aligned}\end{equation}
where we have used scaled time, $\tilde t = t/t_H$. While this is not the same linear ramp-plateau that one obtains in a microcanonical ensemble and is more familiar from the literature, the long-time behaviour of the observables still has a monotonic growth followed by a plateau, \autoref{fig.massive-massless}.

The initial monotonic growth comes from the contributions of the correlations between the energy levels that are separated by a few mean level spacings, $\omega \gtrsim \Delta$. In this regime, $\sin^2(s) \sim 1/2$, can be  averaged over a few level spacings, and the sine-kernel behaves like $1-1/(2 s^2)$. Note that while the ramp in \autoref{fig.massive-massless} only starts at ergodic time, $t_{\rm er}$, as we will discuss in detail in the next section. At even longer times, the regime $s\ll1$ is explored giving rise to the plateau.

\noindent Having explained the ergodic limit in the SYK model and the behaviour of the operators in this limit, we next move on to understand the non-universal deviations of the observables from the RMT-like behaviour at earlier times in the following section.
%%%%%%%%%%%%%%%%%%%%%%%%%%%%%%%%%%%%%%%%%%%%%%%%%%%%%
%%%%%%%%%%%%%%%%%%%%%%%%%%%%%%%%%%%%%%%%%%%%%%%%%%%%%
%%%%%%%%%%%%%%%%%%%%%%%%%%%%%%%%%%%%%%%%%%%%%%%%%%%%%

\section{Non-ergodic modes}\label{sec.massive} 
\begin{figure}[t!]
\begin{subfigure}{.5\textwidth}
  \centering
  \includegraphics[width=.95\linewidth]{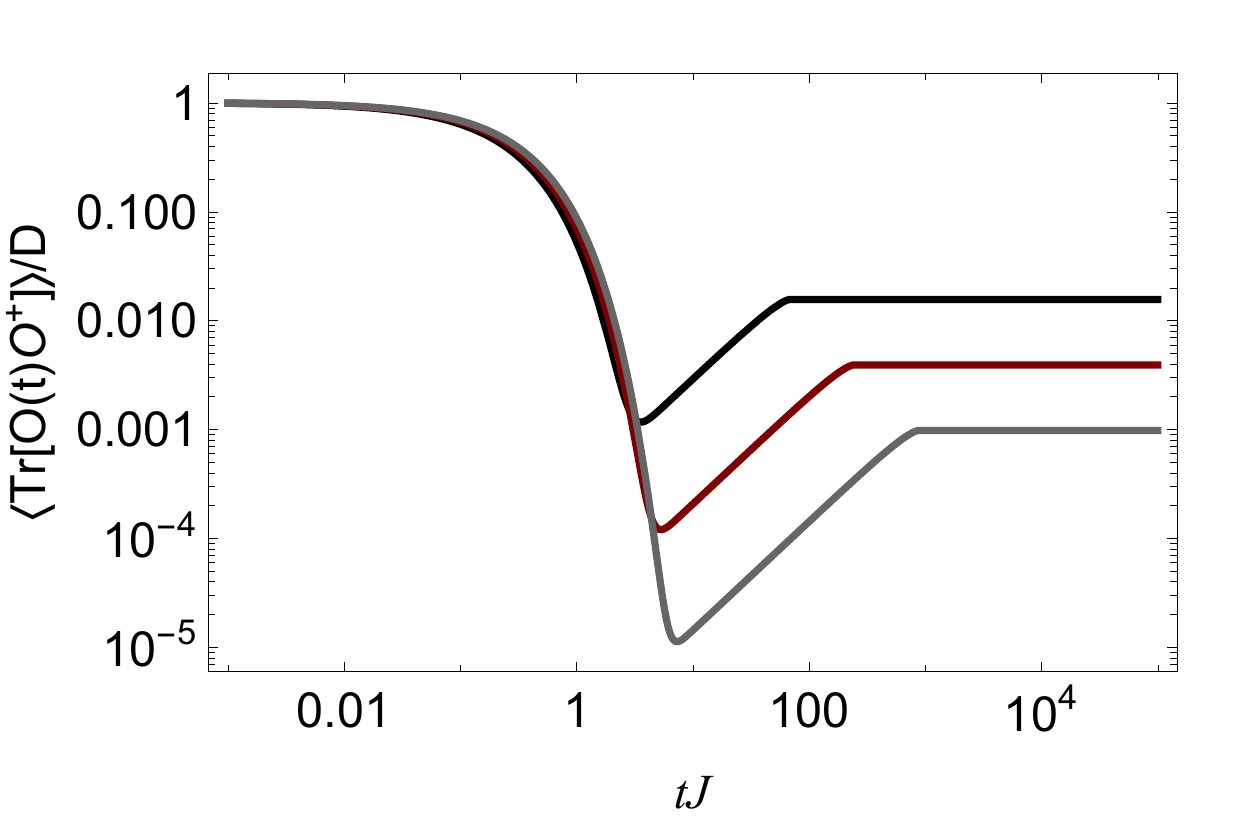}
  \caption{}
  \label{fig:sfig5a}
\end{subfigure}%
\begin{subfigure}{.5\textwidth}
  \centering
  \includegraphics[width=.95\linewidth]{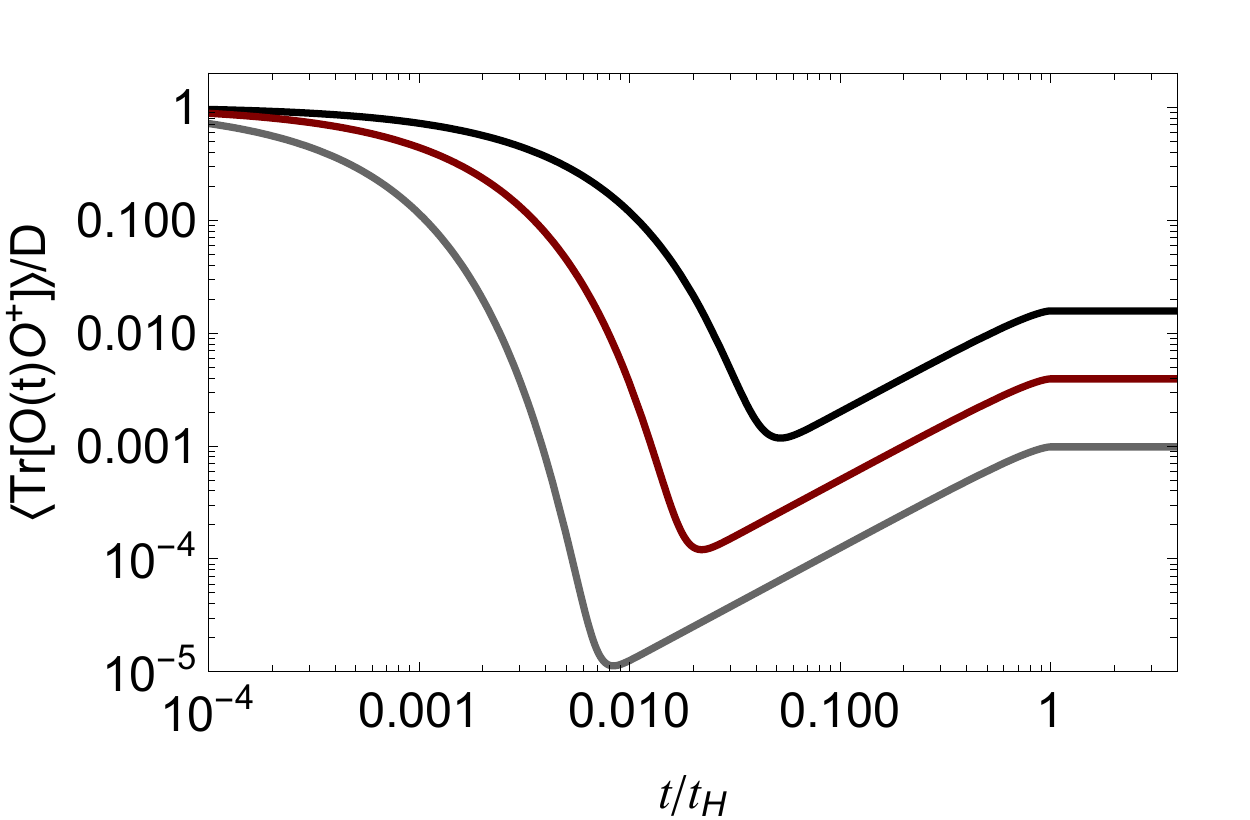}
  \caption{}
  \label{fig:sfig5b}
\end{subfigure}
  \caption{The correlation function $C(t)$ for the two-body hopping operator, $\mathcal O = \psi_i \psi_j$ ($i \neq j$), shown versus (a) physical time $t$ and (b) scaled time $t/t_H$ for three choices of $N$: black curve ($N=14$), red ($N=18$) and grey ($N=22$).}
\label{fig.massive-massless}
\end{figure}

While the SYK model is defined as a disordered Hamiltonian, that is via an average over coupling constants, it is still far from being a random matrix model where every independent entry of the Hamiltonian is a random variable, as emphasised previously in the introduction. Nevertheless after the ergodic time (c.f. \autoref{fig.massive-massless}) every ergodic chaotic quantum system is thought to be well described by random matrix theory.

This includes, as a special case, the quantum chaotic SYK models, i.e.  \eqref{eq.SYKHamiltonian}.  Up to this point, we have concerned ourselves with understanding how the full RMT physics originates in this system and what are its phenomenological effects on the operator correlation functions. However, in a generic quantum chaotic system one expects to see  deviations from the universal RMT behaviour at early enough times. This is no different in the SYK class of models, with the added benefit that we can analytically study the non-ergodic regime as well. While, as we saw, the RMT physics comes from Hilbert space homogeneous modes, the non-universal modes correspond to the projection of the field $\bar A$ in the non-homogeneous directions, $\bar a^{\mu\neq0}$. Previously, we had made an ansatz, \eqref{eq.hha}, to study the saddle point equations arising from \eqref{eq.fullaction}. Fluctuations around the saddle point solution are captured by
\begin{equation}
	\bar A =  y \, \mathds 1 + \sum_{\{\mu_k\}} \bar a^{\mu_k} X^{\mu_k}~.
\end{equation}
These fluctuations contribute to both the terms of the action \eqref{eq.fullaction}. The first quadratic term contributes following additional term to the quadratic action of \eqref{eq.HHA-action},
\begin{equation}
	\sum_{\{\mu_k\}}  (-)^{\frac k2 (k-1)} S(\mu)^{-1} {\rm STr}\left[\bar a ^{\mu_k} \cdot \bar a ^{\mu_k} \right]~.
\end{equation}
We expand the second term in \eqref{eq.fullaction} perturbatively in the fields $\bar a^{\mu_k}$. We do this by writing the matrix $K$ that appears in \eqref{eq.fullaction} as a sum of various components,
\begin{equation*}
	K = K_{0} + K_{\omega} + K_{\mathbb J} + K_{M}~,
\end{equation*}
representing the homogeneous term, frequency $\omega$-dependent terms, sources and non-homogeneous terms, respectively. The last three terms are treated perturbatively in the expansion of the ${\rm STr}[\ln K]$ term. The tracelessness and orthogonality of the $X^{\mu_k}$ matrices imply that only terms that contain even numbers of each of these matrices are non-zero. Thus the action for these modes is even and there are no \emph{tadpole diagrams} that contribute perturbatively (see \autoref{app.massive} for more details, especially the discussion preceding \eqref{eq.massive-quad-app}). Therefore, these modes can be treated perturbatively around the saddle point, \eqref{eq.saddle-sol}, as well as the entire $\sigma$-model manifold. This provides a self-consistent justification of the ansatz. 

Following the detailed analysis of these massive modes presented in \autoref{app.massive}, one finds that for generic observables the two point function is given by,
\begin{align}
	C(t) &= C_{\rm RMT}(t) + \tilde C(t) \nonumber \\
	& = C_{\rm RMT}(t) +  \frac1{D} \, \sum_{\{\mu_{k}\}} f\big(\epsilon_k(0) \, t\big) \, {\rm Tr}\left[O X^{\mu_{k}}\right] {\rm Tr}\left[O^\dagger X^{\mu_{k}}\right]~, \label{eq.Ct-intermediate-main}\\[5pt]
	&f(x) = 1 - \frac 2 \pi \int_0^{x} K_0(y)dy = 1- x \left[ K_0(x) L_{-1}(x) + K_1(x) L_{0}(x) \right],
\end{align}
where, $C_{\rm RMT}(t)$ is given by \eqref{eq.CRMT}. The function, $f(x)$, decays exponentially for $x\gg1$. Therefore, individual terms in \eqref{eq.Ct-intermediate-main} decay at a rate determined by the mass, $\epsilon_k(0)$. The sum in the above expression runs over all the basis elements of the $D\times D$ Hermitian matrices, \eqref{eq.basis-full}. A given typical light operator has non-zero projection only over a few of these basis elements. The ergodic time that marks the transition between the universal late time physics and the non-universal early time physics corresponds to the minimum of the $C(t)$. Quite importantly, it depends on the operator of interest. For a $k$-local operator, $\mathcal O = X_\nu$ with $k = |\nu|$,
\begin{equation}
	t_{\rm er} \sim \frac{N^{3/2}}{J k}~,
\end{equation}
as long as $|N/2-k|\gg1$. In the \autoref{fig.massive-massless}, we have plotted $C(t)$ for the operator $\mathcal O = \psi_i \psi_j$ $(i\neq j)$. These results complements the results of known literature, \cite{Garcia-Garcia:2016mno, Altland:2017eao, Garcia-Garcia:2018ruf, Khramtsov:2020bvs}, where the ergodic time for the spectral form factor was studied. Numerical studies of the ergodic time for operator correlation functions is beyond the scope of the present work and we leave it for future investigations, \cite{Altland.numerics}.
 
%%%%%%%%%%%%%%%%%%%%%%%%%%%%%%%%%%%%%%%%%%%%%%%%%%%%%
%%%%%%%%%%%%%%%%%%%%%%%%%%%%%%%%%%%%%%%%%%%%%%%%%%%%%
%%%%%%%%%%%%%%%%%%%%%%%%%%%%%%%%%%%%%%%%%%%%%%%%%%%%%

\section{Summary \& Discussion}\label{sec.summary}
Over the years, random-matrix physics has been thought to be an important ingredient in understanding the quantum physics of black holes. In recent years this connection has been reinforced, with the surprising realisation that semi-classical gravity itself appears to capture moments of the probability distribution governing this random-matrix like behaviour \cite{Saad:2018bqo,Pollack:2020gfa,Belin:2020hea,Belin:2020jxr,Altland:2020ccq,BdBNS}. In the present work we extended our understanding of the emergence of random matrix behaviour beyond the study of spectral indicators, in particular in operator correlation functions.  Using the causal symmetry breaking approach of \cite{Altland:2020ccq}, we established that at late times two point functions of typical non-extensive operators demonstrate universality in their time dependence, succinctly captured in the operator sine kernel, \eqref{eq.final-sigma}. While the EFT description of the ergodic behaviour -- post Thouless time -- should apply more widely to ergodic theories (e.g. also to higher-dimensional holographic field theories), the approach to and exact value of the Thouless time are not universal. As an example of a holographic quantum system over which we have microscopic control, the SYK model provides a valuable case study that allows us to go further. Indeed we systematically derived the deviations from the universal ergodic behaviour at earlier times. These deviations depend on the theory under consideration, as well as on the operators under study. The techniques discussed in this work hold as long as the operator sources can be treated perturbatively as the analysis of the saddle point and the fluctuations around it is independent of the sources. As we have shown the deviations are controlled by a set of massive modes, inhomogeneous in the Hilbert space, and whose mass is directly related to the amount of inhomogeneity. The technology developed in our work to study the dominance of homogeneous solutions as well as the non-homogeneous mode that are non-universal is generalisable to other strongly interacting Hamiltonians. However, the exact implementation of this technology to more complicated theories is an interesting future challenge. Lastly, it is important to note that the result \eqref{eq.final-sigma.tracefull} is consistent with the prediction of ETH for the two point function: the off-diagonal contribution is suppressed by an $e^{-S}$ factor with respect to the diagonal contribution. However, we emphasise once more that this result encapsulates the information of ETH as well as spectral statistics in it. Comparing our results directly with other proposals of how quantum chaotic physics makes a contribution to gravity is therefore subtle, and we now  undertake a somewhat detailed comparison to other related results that have appeared in the literature, before moving on to a brief outlook of interesting open issues and future directions.

\subsection*{Comparison with known results}
The problem of understanding and characterising the onset of random matrix behaviour in many-body systems was addressed numerically in \cite{Gharibyan:2018jrp}. In the Jackiw-Teitelboim theory, \cite{Teitelboim:1983ux, Jackiw:1984je}, or dilaton-gravity in two dimensions, the emergence of RMT has been studied in \cite{Saad:2018bqo, Saad:2019lba}. In these studies it was argued that  $t_{\rm er} \sim \log(S)$ for the SYK model, where $S$ is the entropy. This argument is supported by explicitly constructing states that contribute to a linear growth of the spectral form factor. However, in \cite{Altland:2017eao}, as well as in the present work, it was demonstrated that the existence of non-universal modes masks this universal linear growth, making it not apparent in observables. Consequently, the linear ramp becomes detectable only after the last/lightest of these non-universal modes have decayed completely. The time scales corresponding to this transition depend on the operators under consideration. For example, $t_{\rm er} \sim \sqrt S \log(S)$ for the spectral form factor and $t_{\rm er}\sim S^{3/2}$ for the kind of $k$-local operators considered in this work. In the double-scaled SYK model, this also agrees with the analytic studies of \cite{Khramtsov:2020bvs}, where the more conventional collective field, or the ``$G,\Sigma$" approach, is used to arrive at the same time scales. Comparisons with numerics in \cite{Garcia-Garcia:2016mno, Garcia-Garcia:2018ruf} lend support to this result. Our results are also consistent with \cite{Kos:2017zjh, Chan:2018dzt}. Numerical studies for the operator correlation functions will be performed in \cite{Altland.numerics}. Having addressed some subtle and perhaps technical issues that have arisen in our work and the literature, we would now like to discuss some of the more qualitative issues implied by the operator statistics derived in this paper.

In a recent work, \cite{Pollack:2020gfa}, it has been proposed that the universal late time behaviour of  observables in semiclassical gravity can be understood to arise from the Haar-averaging over states in a microcanonical energy window. As it turns out this proposal is quite similar to the physics dictated by our EFT, in particular in the case in which one projects the observables in a microcanonical window as described in \autoref{sec.op-res}. Our results imply the picture that wormhole-like contributions in quantum gravity arise from quantum chaos via Haar averaging over eigenstates, but crucially the detailed analysis we presented demonstrates that observables receive contributions from {\it both} the spectral statistics as well as the Haar-averaging over the eigenstates.

Finally, let us comment that recently the reference \cite{Richter:2020bkf} addressed a similar question of finding RMT statistics within physical theories by looking for the signatures of random-matrix behaviour at small frequencies (equivalently, at late times) in operator correlation functions. Our approach can also be used to analytically address the questions studied in this work.

\subsection*{Outlook and generalisations}
The question of generalising the picture we have presented in this work to higher dimensions for theories of physical interest, such as ${\cal N}=4$ SYM remains a non-trivial one. We hope to address this question in  future work. While the optimistic vision would be to use our techniques to understand thermalisation in  $\mathcal N=4$ SYM to understand the questions about apparent thermalisation in high energy states as well as the corresponding holographic statements in the theory of gravity, a more prudent approach might be to apply them to the next simplest example on the ladder of complexity: the AdS$_3$/CFT$_2$ correspondence.
Combined with the results of \cite{Altland:2020ccq}, especially those pertaining to the minimal string theory discussed in that paper, we have a powerful tool promising to unravel the holographic duality with a finer scrutiny at late times.

In relation to the SYK model itself, a relevant question that is studied in \cite{Sedrakyan:2020oip} concerns the relation between the more widely familiar collective field treatment of the SYK model (originally developed by \cite{1999PhRvB..59.5341P, Sachdev:2010um, Kitaev-talks:2015}) to the $\sigma$-model approach discussed in the present work. A unified approach will be quite useful to develop a coherent picture. The recent work, \cite{Goel:2021wim}, proposes a string theoretic origin of the SYK model. It would be interesting to compare the matrix model described in their work with the emergent RMT discussed in our present work.

 It would also be very interesting as well as important to study simple models akin to the SYK model which lie in other random matrix universality classes. Some of these can be achieved within the SYK model itself with different number of Majorana fermions, \cite{Cotler:2016fpe}. We hope to address these directions of research in the  future. How a chaotic quantum system approaches the behaviour of a full dim${\cal H}$ sized random matrix is an extremely important question in general and in particular in the context of holography, where such an understanding promises to microscopically resolve the unitary late time behaviour of black holes. It appears thus of crucial importance to obtain a more complete understanding of causal symmetry breaking in the bulk gravity, the minimal string example worked out in \cite{Altland:2020ccq} being a low-dimensional prototype.

%%%%%%%%%%%%%%%%%%%%%%%%%%%%%%%%%%%%%%%%%%%%%%%%%%%%%
%%%%%%%%%%%%%%%%%%%%%%%%%%%%%%%%%%%%%%%%%%%%%%%%%%%%%
%%%%%%%%%%%%%%%%%%%%%%%%%%%%%%%%%%%%%%%%%%%%%%%%%%%%%

\section*{Acknowledgments}
We would like to thank Alexandre Belin, John Chalker, Jan de Boer, David Huse, Steven Shenker, Jacobus Verbaarschot, Herman Verlinde for enlightening discussions. This work has been supported in part by the Fonds National Suisse de la Recherche Scientifique (Schweizerischer Nationalfonds zur F\"orderung der wissenschaftlichen Forschung) through Project Grants 200020\_ 182513, the NCCR 51NF40-141869 The Mathematics of Physics (SwissMAP), and by the DFG Collaborative Research Center (CRC) 183 Project No. 277101999 - project A03.

\pagebreak
\appendix
\section{Saddle point analysis}\label{app.saddle}
\begin{figure}[htbp]
\begin{subfigure}{.5\textwidth}
  \centering
  \includegraphics[width=.95\linewidth]{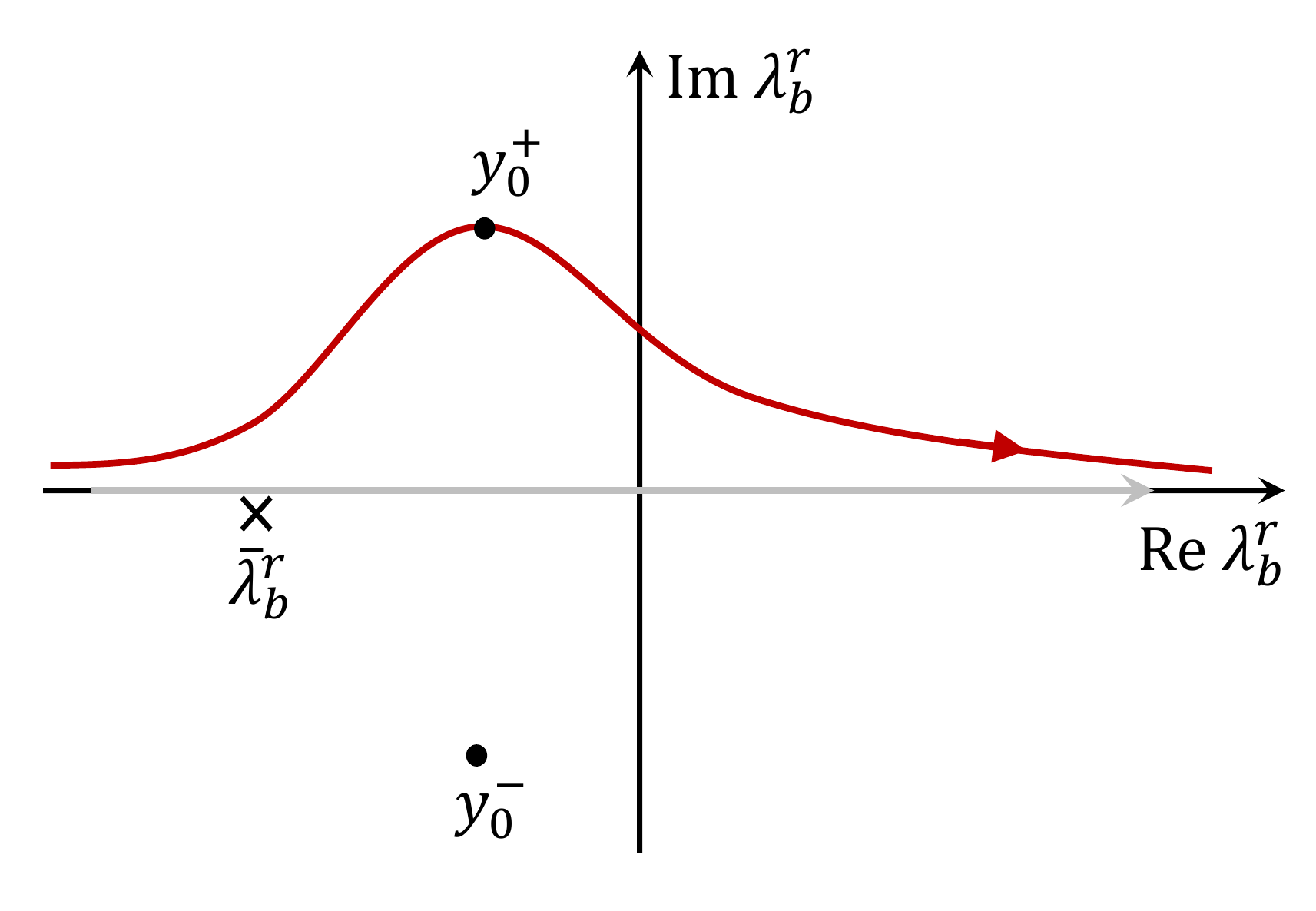}
  \caption{}
  \label{fig.contour-deform-adv}
\end{subfigure}%
\begin{subfigure}{.5\textwidth}
  \centering
  \includegraphics[width=.95\linewidth]{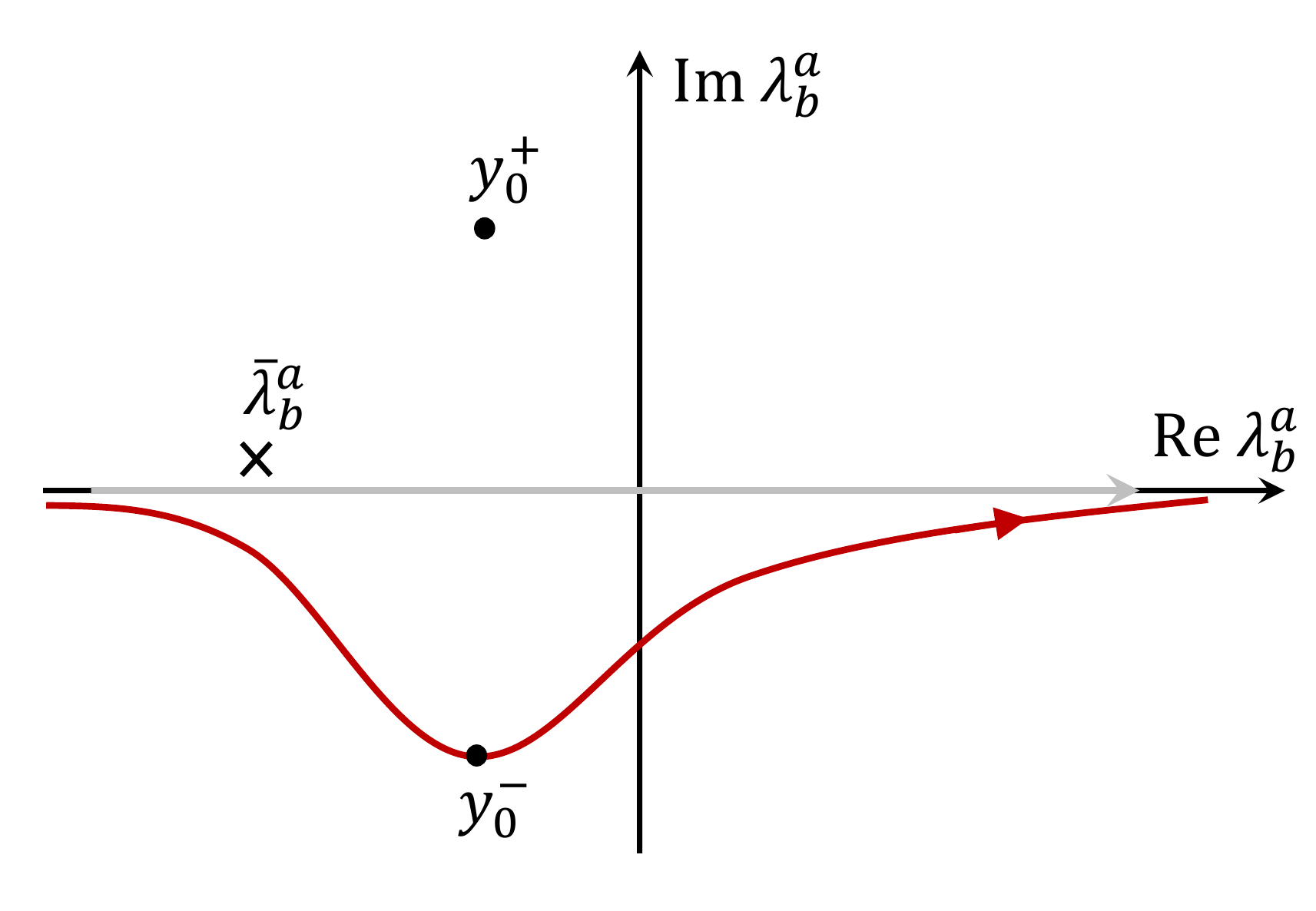}
  \caption{}
  \label{fig.contour-deform-ret}
\end{subfigure}
  \caption{In the boson-boson block, only one of the saddle points $y_0^\pm$ can be reached by a deformation of the 
	original integration contour ${\rm Im}(\lambda_b^{r/a}) = 0$, denoted by the grey line, without crossing the pole. In the above figures, the poles are denoted by black crosses, and the saddle points by black points. Figure (a) demonstrates the deformation of the contour in the retarded space; the figure (b) demonstrates the contour deformation for the advanced space.}
\label{fig.contour-deform}
\end{figure}

The discussion presented in this appendix reproduces the well known results of standard references, \cite{efetov-book, haake-book, Verbaarschot_review}, for the reference of the reader. The path integral after taking the homogenous Hilbert-space ansatz in \eqref{eq.HHA-action} is given by,
\begin{eqnarray}\label{eq.HHA-action-app}
	Z &=& \int \!\! \mathcal D  y \exp\left[ -\frac D2  {\rm STr}(y^2) - {\rm STr} \ln K \right]\\
	K &=& E \; \mathds{1}^{\rm RA} \otimes \mathds{1}^{\rm bf} \otimes \mathds{1}^{\rm Fock} + \gamma  y \otimes \mathds{1}^{\rm Fock},
\end{eqnarray}
where we have dropped the terms that we treat perturbatively in our analysis (the sources and the $\omega\neq0$ terms). The saddle point equation in the large $D$ limit is,
\begin{equation}
	-\frac{ y}\gamma = \left( E \mathds{1}^{\rm RA} \otimes \mathds{1}^{\rm bf} + \gamma  y \right)^{-1}~.
\end{equation}
The solution of this equation is given by,
\begin{equation}
	 y_0 = -\dfrac E{2\gamma} \mathds{1}^{\rm RA} \otimes \mathds{1}^{\rm bf} + i \tilde\Lambda \sqrt{1-\frac{E^2}{4\gamma^2}}~,
\end{equation}
where, $\tilde\Lambda$ is a diagonal matrix with entries $\pm1$. However, only a few of these solutions can be reached by contour deformations starting from the initial path integral. Let us denote by $\lambda_{b,f}^{a,r}$ the eigenvalues of the $y$-matrix in the boson/fermion and advanced/retarded sectors, respectively. When integrating along $\lambda_{b}^{a,r}$ in the bosonic sector of the $y$-path integral, one encounters
poles at
\begin{equation}
	\bar\lambda^r_b = -\frac1{ \gamma } \left( E + i 0^+ \right), \qquad \bar\lambda^a_b = -\frac1{ \gamma } \left( E - i 0^+ \right)~.
\end{equation}
Note, that only one saddle point, either $y_0^+ = -\frac E{2\gamma} + i \sqrt{1-\frac{E^2}{4\gamma^2}}$ or $y_0^- = -\frac E{2\gamma} - i \sqrt{1-\frac{E^2}{4\gamma^2}}$, can be reached by a contour deformation in the retarded/advanced sectors, respectively (see \autoref{fig.contour-deform}).

\begin{figure}[htbp]
\begin{center}
	\begin{minipage}{.5\textwidth}
  		\centering
 		 \includegraphics[width=.95\linewidth]{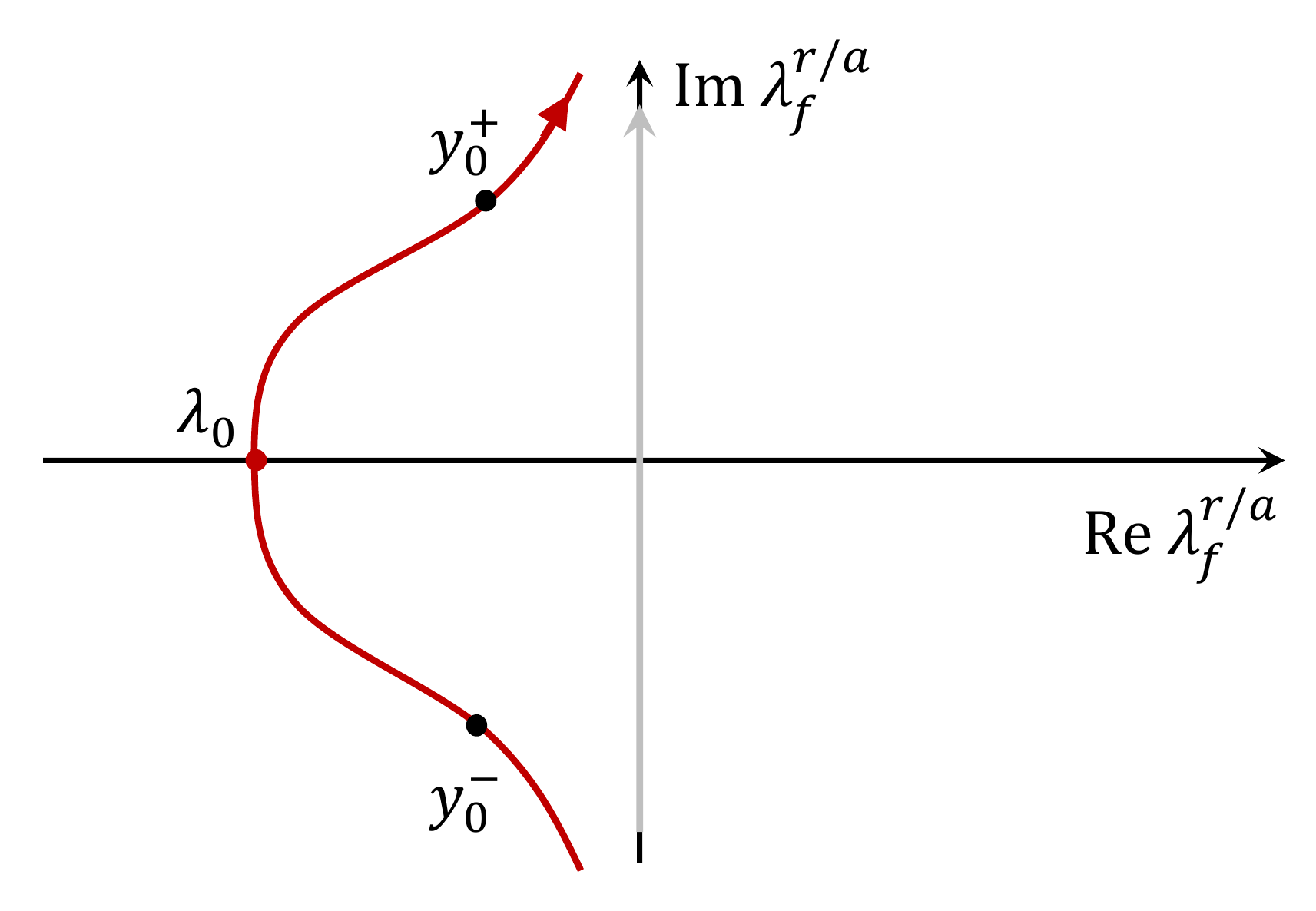}
		  \label{fig.contour-deform-Fermi-adv}
	\end{minipage}%
\caption{For the eigenvalue integrals in the fermionic sector, the contour of stationary phase passes through both the saddle points $y_0^\pm$
	and the zero $\lambda_0$. However, the contribution of one of the saddle points is sub-leading with respect to the other.
    The original integration contour, ${\rm Re}(\lambda_f^{r/a}) = 0$, is denoted by the grey line.}
\label{fig.contour-deform-Fermi}
\end{center}
\end{figure}

On the other hand, the eigenvalues in the `Fermi'-block don't have any poles, but zero at the point $\lambda_0 = -\frac{E}{2\gamma}$. The singularity structure for these integrals is demonstrated in \autoref{fig.contour-deform-Fermi}. It is worth noting here that 
${\rm STr}( P_f y^2 ) = - {\rm tr} (P_f y^2)$, thus the original integration contour runs along the imaginary axis in 
the complex plane of variables $\lambda_f^{r/a}$. 
It can be shown, see monograph \cite{haake-book}, that the leading saddle point corresponds to,
\begin{equation}
	\tilde\Lambda = \sigma_3^{\rm RA} \otimes \mathds{1}^{\rm bf} =: \Lambda
\end{equation}
The other saddle point solution corresponds to, \cite{Andreev:1995},
\begin{equation}
	\tilde\Lambda = \sigma_3^{\rm RA} \otimes \sigma_3^{\rm bf} =: \Lambda_{AA}
\end{equation}
$T_0 \in U(1,1|2)/U(1|1)\otimes U(1|1)$, which transforms the standard saddle point, $\Lambda$ to the sub-leading saddle $\Lambda_{AA}$ is given by,
\begin{equation}\label{eq.weyl-matrix}
	T_0 = \mathds{1}^{\rm RA} \otimes P_b +  \sigma_1^{\rm RA}  \otimes P_f, \qquad T_0 \cdot \Lambda \cdot T_0^{-1} = \Lambda_{AA}
\end{equation}

\section{Perturbation analysis around the saddle points}\label{app.details}
\subsection{\texorpdfstring{The $\sigma$}{sigma}-model with source terms}
We begin our analysis by expanding the trace-log piece of the action, \eqref{eq.HHA-action-app}, containing the terms
\begin{equation}\label{eq.pert-H}\begin{aligned}
K &= K_0 + K_\omega +  h,\qquad
&K_0 = E + \gamma  y,\qquad
&K_\omega = \left(\frac\omega2 + i 0^+\right)\sigma_3^{\rm RA} \otimes \mathds 1^{\rm bf},
\end{aligned}\end{equation}	  
so that
\begin{equation}\label{eq.pert-H-exp}
 {\rm STr} \ln K  = {\rm STr} \ln K_0  + {\rm STr}[K_0^{-1}K_\omega + K_0^{-1}  h] - \frac 12 {\rm STr}[K_0^{-1}  h K_0^{-1}  h] + \ldots~.
\end{equation}
The terms in the above expansion are (1) at most linear in $\omega$, and (2) at most quadratic in sources. In addition, we use
\begin{equation}
	K_0^{-1} = - \frac{ y}\gamma~,
\end{equation}
to write
\begin{equation}
	{\rm STr} \ln K  = {\rm STr} \ln K_0  - {\rm STr}[\frac{ y}\gamma K_\omega +  \frac{ y}\gamma  h] - \frac 12 {\rm STr}[ \frac{ y}\gamma  h  \frac{ y}\gamma  h] + \ldots~.
\end{equation}
The saddle solution of $ y$ is
\begin{equation}
	 y = -\frac E{2\gamma} + \frac i{2\gamma} \Lambda \sqrt{4\gamma^2-E^2}, \quad \Lambda = \sigma_3^{\rm RA} \otimes \mathds 1^{\rm bf} = \tau_3~,
\end{equation}
here we have defined, $\tau_3 = \sigma_3^{\rm RA} \otimes \mathds 1^{\rm bf}$, as the Pauli matrix in the adv./ret. space. The perturbations of the saddle point by $T\in U(1,1|2)/U(1|1)\times U(1|1)$ can be parametrised by,
\begin{equation}
	T yT^{-1} = -\frac E{2\gamma} + i \frac{\pi \gamma}D \rho(E) Q, \quad Q = T\Lambda T^{-1}
\end{equation}
$T$ matrix is perturbatively written as
\begin{equation}
	T = \mathds{1} - W + \frac 12 W^2 - \frac16 W^3 + \frac1{24} W^4 + \cdots~,
\end{equation}
$W$ is the pion field,
\begin{equation}\label{eq.WB.def}
	W = -\begin{pmatrix}
		0 & B\\
		\tilde B & 0
	\end{pmatrix}~.
\end{equation}

\subsection{Gaussian integration around the standard saddle point}
\begin{figure}[!htbp]
  \centering
    \includegraphics[width=0.45\textwidth]{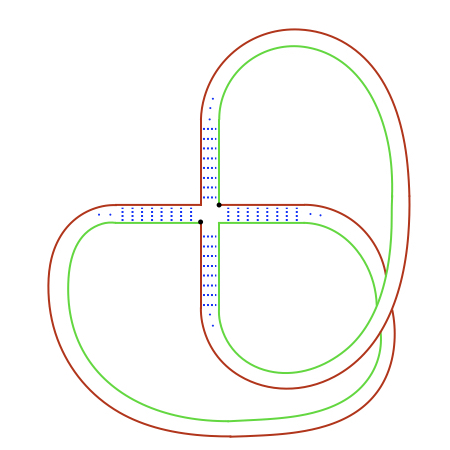}
  \caption{The contribution to the operator correlation function arises from the non-planar diagram.}
  \label{fig.subleading.app}
\end{figure}

In this case the perturbations are parametrised by
\begin{equation}
T \Lambda T^{-1} = \tau_3 + 2 \tau_3 W + 2 \tau_3 W^2 + \frac43 \tau_3 W^3 + \frac23 \tau_3 W^4 + \cdots~,
\end{equation}
	By defining the diagonal matrix
	\begin{equation}
	{\rm diag}\{e_1,e_3,e_2,e_4\} = \pi \rho(E) (\omega + i 0^+) \sigma_3^{\rm RA},
	\end{equation}
	where the bosonic eigenvalue $e_1$ and the fermionic one, $e_3$, form the retarded sector, 
	one can use it to write the individual terms as 
\begin{eqnarray}
	{\rm STr}\left[\frac{ y}\gamma K_\omega \right] &=& -\frac E{2 \, \sqrt{4\gamma^2-E^2}} {\rm STr}\left[ \hat e \right]+\frac i {2} {\rm STr}\left[Q\hat e\right],\\
								     &=& -\frac E{2 \, \sqrt{4\gamma^2-E^2}} {\rm STr}\left[ \hat e \right]+\frac i {2} {\rm STr}\left[T\Lambda T^{-1}\hat e\right],\nonumber\\
								     &=&\frac i {2} \left(e_1-e_3-e_2+e_4\right){-\frac E{2 \sqrt{4\gamma^2-E^2}}\left(e_1-e_3+e_2-e_4\right)}\nonumber\\
								     &+&i~ {\rm STr}\left[ B\tilde B \begin{pmatrix}e_1&0\\0&e_3\end{pmatrix} - \tilde B B \begin{pmatrix}e_2&0\\0&e_4\end{pmatrix} \right] + \nonumber \\
								     &+&\frac {2i} {3} {\rm STr} \left[\left(B\tilde B\right)^2 \begin{pmatrix}e_1&0\\0&e_3\end{pmatrix}- \left(\tilde B B\right)^2 \begin{pmatrix}e_2&0\\0&e_4\end{pmatrix}\right] + \ldots~.
\end{eqnarray}
The first three terms are the kinetic terms for our \emph{perturbation} theory and the last term as well as the higher-order terms that are denoted by the ellipsis can be treated perturbatively. However, we find that these higher order terms are not relevant due to non-renormalisation theorems that are applicable for the GUE. It is useful to write the propagator for the $B, \tilde B$ fields as (using $e_3=e_1, e_4=e_2$)
\begin{equation}\label{eq.BBtprop}\begin{aligned}
	{\rm STr}[B\cdot X] \, {\rm STr}[\tilde B \cdot Y] &= \frac{i}{e_1-e_2} {\rm STr}[X Y]\\
	{\rm STr}[B \cdot X \cdot \tilde B\cdot Y] &=\frac{i}{e_1-e_2} {\rm STr}[X] {\rm STr}[ Y]\\
\end{aligned}\quad,\end{equation}
where, $X,Y$ are arbitrary supermatrices. Subsequently, we expand the source terms to quartic order in $W$.\footnote{We are using the matrices $B$ and $\tilde B$ instead of $W$ because they are more convenient to work with. If $\mathcal O^\dagger = \mathcal O$ then we could continue to use $W$ with equal ease.} However, to avoid clutter we explicitly write only the terms that are non-zero using the above contraction rules.
\begin{equation} \label{eq.2trace.b4contraction-app}
	\partial_{h_+}\, {\rm STr}\left[\frac{ y}\gamma  h\right]\, \partial_{h_-}\, {\rm STr}\left[\frac{ y}\gamma  h\right]
									=-\frac {4\pi^2} {D^2} \, \rho^2(E) \, {\rm STr} \left[\tilde B \cdot P_b \right] {\rm STr} \left[ B\cdot P_b \right]{\rm Tr}[\mathcal O] {\rm Tr}[\mathcal O^\dagger] + \cdots 
\end{equation}
and,
\begin{align}
	\frac12\partial_{h_+}\partial_{h_-} \,{\rm STr}\left[\frac{ y}\gamma  h\frac{ y}\gamma  h\right]
											  &= {- \left( -\frac E {2\gamma^2} \right)^2} {\rm Tr} \left[ \mathcal O \mathcal O^\dagger \right]\nonumber \\
																			    &\hspace{-1cm}+\frac12 \left(\frac i {2\gamma^2} \sqrt{4\gamma^2-E^2} \right)^2 {\rm STr} \begin{bmatrix} 2P_b + 8 \tilde B P_b \tilde B B P_b B \end{bmatrix} \times {\rm Tr}[\mathcal O \mathcal O^\dagger] + \cdots~.
\end{align}
So the path integral that we are interested in computing is given by
\begin{equation}\begin{aligned}\label{eq.path.integral.interest.pert.std.app}
	\partial_{h_+}\partial_{h_-}Z|_{h_\pm=0} &= \int  \mathcal DT  \, \exp\left[  {i(e_1-e_2)}{\rm STr}(B\tilde B )\right] \times \left[\partial_{h_+} S_{\rm src} \, \partial_{h_-} S_{\rm src}- \partial_{h_+}\partial_{h_-}S_{\rm src}\right]_{h_\pm=0}~,
\end{aligned}\end{equation}
where, $S_{\rm source}$ is the shorthand for all the source, $ h$,  dependent terms. Note that only the term linear in $h$ contributes to the term that involves a single derivative with respect to $h$ and the term quadratic in $h$ to the second term. However, since we are working with traceless operators \eqref{eq.2trace.b4contraction-app} is zero. The vertex in \eqref{eq.path.integral.interest.pert.std.app} can be diagrammatically represented in the double-line notation as shown in \autoref{fig.subleading.app}. The coloured fat-line propagators demonstrate matrix fields $B,\tilde B$ respectively in the double line notation. These $B,\tilde B$ fields have one index each in advanced and the retarded sector as is apparent from its definition in \eqref{eq.WB.def}, which is demonstrated by different colours of the double lines. The fields $B, \tilde B$ have a concealed trace over the Fock-space, generated by the integration over the Hamiltonian (see \eqref{eq.Z-disordered-avg-Hubb-Strat}), which is represented by the blue dotted lines in this diagram. The insertion of projectors in the vertex is demonstrated by the black dots. The projectors in the definition of the source, \eqref{eq.src}, define how they change the causalities in this diagram.

Evaluating the path integral one gets,
\begin{equation}\label{eq.ans.std.app}
	- \frac{4\pi^2}{(e_1-e_2)^2} \, \rho^2(E) \, \frac{{\rm Tr}\left[ \mathcal O \mathcal O^\dagger \right] }{ D^2}  + \frac1 {\gamma^2} {\rm Tr}\left[ \mathcal O \mathcal O^\dagger \right]~.
\end{equation}

\subsection{Gaussian integration around Andreev-Altshuler saddle point}
In general, our analysis remains the same as above. However, there are some important qualitative differences between the two path integrals. Around the subleading saddle point, that can be obtained by a Weyl transformation of the standard saddle point as discussed in \eqref{eq.weyl-matrix}, the kinetic term for the Grassmann variables, $\mu,\bar\mu,\nu,\bar\nu$ vanishes. Thus, for the Grassmann integrals to be non-vanishing, all four factors of the Grassmann variables should arise from the pre-exponent terms. This means that the first non-trivial contribution to the path integral comes from terms that are quartic in the pion fields $W$ in the pre-exponential factors. Explicitly,
\begin{equation}\begin{aligned}
	{\rm STr}\left[\frac{ y}\gamma K_\omega\right] &= i \left(e_1-e_2\right) + \frac {i(e_1-e_2)}{2} {\rm STr}\left[\sigma_3 B\tilde B+ \sigma_3 \tilde B B \right] + \cdots~,
\end{aligned}\end{equation}
where, once again the higher order terms don't contribute. Of all the terms arising in the expansion of the source term only the following terms contribute,
\begin{equation}
	\frac12\partial_{h_+}\partial_{h_-} \,{\rm STr}\left[\frac{ y}\gamma  h\frac{ y}\gamma  h\right] \approx -{4\pi^2} \rho^2(E)\, \frac{{\rm Tr}[\mathcal O\mathcal O^\dagger]}{D^2} \times {\rm STr}\left[B \tilde B P_b \tilde B B P_b\right]~.
\end{equation}
Note that diagrammatically this vertex corresponds to the same Feynman diagram as given in \autoref{fig.subleading.app}, however, this time the fields are expanded around the Andreev-Altshuler saddle point. All the above terms contribute equally to the path integral giving a total contribution,
\begin{equation}
	-4 \pi^2 \, \rho^2(E)\, \frac { e^{i  (e_1-e_2) }}{(e_1-e_2)^2} \times \frac{{\rm Tr}[\mathcal O\mathcal O^\dagger]}{D^2}~.
\end{equation} 
Adding this to the above answer, \eqref{eq.ans.std.app}, and taking the real part we obtain:
\begin{equation}\begin{aligned}
	R^\pm(E,\omega) &= 2 \pi^2  \rho^2(E) \left( \pi \delta(s) - \frac {\sin^2(s)} {s^2} \right)\frac{{\rm Tr}\left[\mathcal O \mathcal O^\dagger \right]}{D^2}
	+ \frac{{\rm Tr}\left[\mathcal O \mathcal O^\dagger \right]}{\gamma^2}.
\end{aligned}\end{equation}
To this answer, we still need to add the contribution of the correlation function involving identical causalities.

\subsection{Correlation function with two advanced Green's function}\label{app.plusplus}
The correlation function that appears in the second term on the RHS of \eqref{eq:resolvent} can be defined as a path integral analogous to the first term,
\begin{equation}
{\rm Tr}{}_{\mathcal H}\left[G^+(E + \tfrac \omega 2){\mathcal O} G^+ (E - \tfrac \omega 2) {\mathcal O}^\dagger\right] = \left.{\partial_{h_+} \partial_{h_-}Z_{\ddag}[ h]}\right|_{h=0} 
\end{equation}
where
\begin{align*}
Z_{\ddag}[ h]& = \int \!\! \mathcal D \bar \Psi \mathcal D \Psi \exp\left[ i \bar \Psi \cdot \left( z_{\ddag} - H +  h \right)\cdot \Psi \right] \\
z_{\ddag} &= \frac\omega2 \sigma_3^{\rm RR} \otimes \mathds 1^{\rm bf} \otimes \mathds 1^{\rm Fock} + (E+i0^+)\mathds 1^{\rm RR} \otimes \mathds 1^{\rm bf} \otimes \mathds 1^{\rm Fock}
\end{align*}
Note that in this case, the supervectors $\bar \Psi, \Psi$ comprise two copies of retarded space as opposed to a copy of advanced and retarded space each. It is not hard to see, following the discussion of \autoref{app.saddle}, that the saddle point for the path integral that arises in this case is
\begin{equation}
	 y_{\ddag} = \left(-\frac E{2\gamma} + i \sqrt{1-\frac{E^2}{4\gamma^2}}\right) \mathds 1~.
\end{equation}
The expansion in small $\omega$ to linear order and in sources to quadratic order is given by,
\begin{equation}\begin{aligned}
	{\rm STr}\ln K  &= {\rm STr} \ln K_0  - \frac 1 \gamma\left(-\frac E{2\gamma} + i \sqrt{1-\frac{E^2}{4\gamma^2}}\right) {\rm STr}[K_\omega +  h] \\
					     & \qquad - \frac1{2\gamma^2}\left(-\frac E{2\gamma} + i \sqrt{1-\frac{E^2}{4\gamma^2}}\right)^2 {\rm STr}\left[ h^2\right] + \cdots\\
					     & \approx  \frac{h_+ h_-}{\gamma^2}\left(-\frac E{2\gamma} + i \sqrt{1-\frac{E^2}{4\gamma^2}}\right)^2 \, {\rm Tr}[\mathcal O \mathcal O^\dagger]~.
\end{aligned}\end{equation}
Taking the $h$-derivative twice, we get
\begin{equation}\label{eq.res-plusplus}
{\rm Re} \,{\rm Tr}{}_{\mathcal H}\left[G^+(E + \tfrac \omega 2){\mathcal O} G^+ (E - \tfrac \omega 2) {\mathcal O}^\dagger\right] =
{\rm Re} \, \frac1{\gamma^2}\left(-\frac E{2\gamma} + i \sqrt{1-\frac{E^2}{4\gamma^2}}\right)^2 \, {\rm Tr}[\mathcal O \mathcal O^\dagger] =: R^{\ddag}(E,\omega)~.
\end{equation}

\section{Perturbation analysis of the massive modes}\label{app.massive}
Let's go back to equation \eqref{eq.fullaction},
\begin{equation}\label{eq.fullaction-again}
	Z[ h] \sim \int \!\! \mathcal D \bar a^\mu \; \exp\left[ -\frac D2 \sum_{\mu} s(\mu)S(\mu)^{-1} \; {\rm STr} \left[ \bar a^\mu \cdot \bar a^\mu \right] - {\rm STr}[\ln(K)]\right]
\end{equation}
where,
\begin{equation*}
K =  z +  h + \gamma  y\otimes \mathds 1^{\rm Fock} + \gamma  \sum_{k=1}^N\sum_{\substack{\mu_k\in\\\left\{{N\choose k}\right\}}}\bar a^{\mu_k} X^{\mu_k}~,
\end{equation*}
where the quadratic term can we rewritten more explicitly as,
\begin{equation}\label{eq.quad-mass}
	{\rm STr}\left[  y^2 \right] + \sum_{k=1}^N\sum_{\{\mu_k\}\in\{{N\choose k}\}} (-)^{\frac k2 (k-1)} S(\mu)^{-1} {\rm STr}\left[\bar a ^{\mu_k} \cdot \bar a ^{\mu_k} \right]~.
\end{equation}
Here, we have explicitly split the `notational' $\mu$-sum in \eqref{eq.fullaction-again} as a sum over cardinality of $|\mu| = k$, as well as, a sum over $N\choose k$ choices of fermions for each $k$. We have also separated out the $k=0$ (HHA) term from the remaining terms.
We refer to the non-homogeneous modes as the \emph{massive} modes and treat them perturbatively. The following analysis provides a self-consistent justification for this approach as well as the HHA ansatz that we had taken in \autoref{sec.erglimSYK}. We write $K$ as a sum over various pieces,
\begin{align}
	K =& K_{0} + K_{\omega} + K_{\mathbb J} + K_{M} \\
	K_{M} =& \gamma  \sum_{k=1}^N\sum_{\substack{\mu_k\in\\\left\{{N\choose k}\right\}}}\bar a^{\mu_k} X^{\mu_k} \nonumber
\end{align}
where the rest of the definitions are the same as in \eqref{eq.pert-H}. To treat $K_{M}$ perturbatively, we need to assume $K_M \ll K_{0}$. The linear and quadratic terms in the expansion of the logarithm are,
\begin{align}\label{eq.log-exp-mass}
	{\rm STr}\left[ K_{0}^{-1} K_M \right] &= 0 \\
	-\frac12 {\rm STr}\left[  K_{0}^{-1} K_M  K_{0}^{-1} K_M \right] &= -\frac12 \gamma^2 \sum_{k,k'}\sum_{\mu_k,\mu_{k'}}{\rm STr}\left[ \left( {K}_{0}^{-1} \bar a^{\mu_k} {K}_{0}^{-1} \bar a^{\mu_{k'}}\right) \otimes \left( X^{\mu_k}X^{\mu_{k'}} \right)\right] \nonumber \\
	&= -\frac D2 \gamma^2 \sum_{k}\sum_{\mu_k} (-)^{k(k-1)/2} \, {\rm STr}\left[ {K}_{0}^{-1} \bar a^{\mu_k} {K}_{0}^{-1} \bar a^{\mu_{k}}\right]
\end{align}
where the linear term is zero because the matrices $X^\mu$ are traceless. In fact, this implies that all odd powers in the logarithmic expansion are zero. The potential for the non-homogeneous fields is even with a saddle point solution at zero. This is a stable saddle point which justifies our perturbative treatment as well as homogeneous Hilbert space ansatz.

Using the identity
$$
	K_0^{-1} = - \frac { y}\gamma~,
$$
the quadratic action is
\begin{align}\label{eq.massive-quad-app}
	&-\frac D2 \left[{\rm STr}\left[  y^2 \right] + \sum_{k=1}^{N} (-)^{\frac k2(k-1)}\sum_{\mu_k} \Bigg(  S(\mu)^{-1} \, {\rm STr}\left[\bar a ^{\mu_k} \cdot \bar a ^{\mu_k} \right] - \, {\rm STr}\left[  y \cdot \bar a ^{\mu_k} \cdot  y \cdot \bar a^{\mu_{k}} \right] \Bigg) \right]~.
\end{align}
It is useful to break the fields $\bar a^{\mu_k}$ into diagonal and off-diagonal components,
\begin{equation}\begin{aligned}
	\bar a^{\mu_k} &= w^{\mu_k} + v^{\mu_k}\\
	&w^{\mu_k} = \begin{pmatrix}
		0&W^{(+-)}\\
		W^{(-+)}&0
	\end{pmatrix}, \text{ and, } v^{\mu_k} = \begin{pmatrix}
		V^{(++)}&0\\
		0&V^{(--)}
	\end{pmatrix}
\end{aligned}\end{equation}
because they don't interact.

In what follows we proceed with the analysis of fluctuations around the standard saddle point. In this case the diagonal part of $\bar a ^{\mu_k}$ commutes with $\Lambda$ while the off-diagonal part anti-commutes. Thus we have an action in terms of these fields
\begin{equation}\begin{aligned}
	&-\frac D2  \sum_k (-)^{\frac k2(k-1)}\sum_{\mu_k} \Bigg( S(\mu)^{-1}  {\rm STr}\left[w ^{\mu_k} \cdot w ^{\mu_k}  \right] -  \, {\rm STr}\left[  y_0 \,  y_0^* \, w^{\mu_k} \, w^{\mu_{k}} \right] \Bigg) \\
	&-\frac D2  \sum_k (-)^{\frac k2(k-1)}\sum_{\mu_k} \Bigg( S(\mu)^{-1} {\rm STr}\left[v ^{\mu_k} \cdot v ^{\mu_k}  \right] -  \, {\rm STr}\left[  y_0 \,  y_0 \, v^{\mu_k} \, v^{\mu_{k}} \right] \Bigg)~,
\end{aligned}\end{equation}
where, $ y_0^* = -\frac E{2\gamma} \mathds 1 - i \Lambda \sqrt{1-\frac{E^2}{4\gamma^2}}$ and $y_0 y_0^* = \mathds 1$. We simplify the action further to get,
\begin{equation}\label{eq.quadratic-massive}\begin{aligned}
	&-\frac D2 \sum_k\sum_{\mu_k} (-)^{\frac{k(k-1)}2} \Bigg( \left(S(\mu)^{-1}-1\right){\rm STr}\left[w ^{\mu_k} \cdot w ^{\mu_k}  \right] + {\rm STr}\left[ \left( S(\mu)^{-1} \mathds 1-  y_0^2 \right) \, v^{\mu_k} \, v^{\mu_{k}} \right] \Bigg)~.
\end{aligned}\end{equation}
The quadratic action for the massive mode additionally receives contributions from the $\omega \neq 0$ corrections,
\begin{align}
	\frac {D}{2\pi^2 \gamma \rho(E)} \sum_{k,\mu_k} (-)^{\frac k2(k-1)}{\rm STr}\left[  y_0 \cdot \hat e \cdot \left(\left(w^{\mu_k}\right)^2+  y_0^2 \cdot \left(v^{\mu_k}\right)^2\right) \right]. \label{eq.wvomega} 
\end{align}
The kinetic action for the massive modes is sum of \eqref{eq.wvomega} and \eqref{eq.quadratic-massive},
\begin{equation}\begin{aligned}\label{eq.massiveQuadAct}
	&-\frac D2\sum_{k,\mu_k} (-)^{\frac{k(k-1)}2} \left( {\rm STr}\left[ \left(S(\mu)^{-1}-1\right)w ^{\mu_k} \cdot w ^{\mu_k} - \frac{ y_0 \cdot \hat e}{\pi^2 \gamma \rho(E)}\cdot w ^{\mu_k} \cdot w ^{\mu_k} \right] \right. \\
	&\left.\hspace{4cm}+ {\rm STr}\left[ \left(S(\mu)^{-1}\mathds 1- y_0^2\right)\cdot v ^{\mu_k} \cdot v ^{\mu_k} - \frac{ y_0 \cdot \hat e \cdot  y_0^2}{\pi^2 \gamma \rho(E) }\cdot v ^{\mu_k} \cdot v ^{\mu_k} \right]\right)~.
\end{aligned}\end{equation}

Note that the quadratic integral for massive modes, \eqref{eq.massiveQuadAct}, are proportional to ${\rm STr}\left[ W^{+-} W^{-+} \right]$, ${\rm STr}\left[ {V^{++}}^2 \right]$ and ${\rm STr}\left[ {V^{--}}^2 \right]$. It is easy (although cumbersome) to check that the terms that are quadratic in source, $h$, don't have non-zero massive propagators and therefore don't contribute to the order that we are interested in. The terms linear in $h$ are, 
\begin{subequations}\begin{align}
	&\gamma {\rm STr}\left[ K_0^{-1} \cdot   h \cdot K_0^{-1} \cdot \bar a^{\mu_k}\otimes X^{\mu_k} \right] \nonumber\\
	&\hspace{1cm}= \frac1{ \gamma } \Big( h_+ \, {\rm STr}\left[P_b W_{\mu_k}^{-+}\right] {\rm Tr}\left[ O X^{\mu_k}\right] - h_- {\rm STr}\left[P_b W_{\mu_k}^{+-}\right] {\rm Tr}\left[ O^\dagger X^{\mu_k} \right]\Big)~,\\[20pt]%%%%%%%%%%%%%%%%%%%%%%%%%%%%%%%%%%%%%%%%%%%%%%%%%%%%%%%%%%%%%%%%%%
	& -\gamma^2 {\rm STr}\left[ K_0^{-1} \cdot   h\cdot    K_0^{-1} \left(\bar a^{\mu_k}\otimes X^{\mu_k}\right) K_0^{-1} \left(\bar a^{\mu_{k'}}\otimes X^{\mu_{k'}}\right)\right] \nonumber\\
	&\quad = -\frac 1\gamma \left( \frac {h_+ \, E}{2\gamma} {\rm STr}\left[P_b\cdot W^{-+}_{\mu_k}\cdot V_{\mu_{k'}}^{++} + P_b \cdot V^{--}_{\mu_k}\cdot W_{\mu_{k'}}^{-+} \right] {\rm Tr} \left[O X^{\mu_k} X^{\mu_{k'}}\right]\right.\nonumber\\
	&\hspace{3.5cm} -  \frac { h_- \,E} {2\gamma} {\rm STr}\left[ P_b \cdot V^{++}_{\mu_k}\cdot W_{\mu_{k'}}^{+-} + P_b \cdot W^{+-}_{\mu_k}\cdot V_{\mu_{k'}}^{--} \right] {\rm Tr} \left[O^\dagger X^{\mu_k} X^{\mu_{k'}}\right] \nonumber \\
	&\hspace{3.5cm} -i h_+ \, \sqrt{1-\frac {E^2}{4\gamma^2}} \ {\rm STr}\left[ P_b \cdot W^{-+}_{\mu_k}\cdot V_{\mu_{k'}}^{++} - P_b \cdot V^{--}_{\mu_k}\cdot W_{\mu_{k'}}^{-+} \right] {\rm Tr} \left[O X^{\mu_k} X^{\mu_{k'}}\right]\nonumber\\
	&\hspace{2.75cm} \left. +  i h_-\,  \sqrt{1-\frac {E^2}{4\gamma^2}} \ {\rm STr}\left[ P_b \cdot V^{++}_{\mu_k}\cdot W_{\mu_{k'}}^{+-} - P_b \cdot W^{+-}_{\mu_k}\cdot V_{\mu_{k'}}^{--} \right] {\rm Tr} \left[O^\dagger X^{\mu_k} X^{\mu_{k'}}\right] \right)~.%%%%%%%%%%%%%%%%%%%%%%%%%%%%%%%%%%%%%%%%%%%%%%%%%%%%%%%%%%%%%%%%%%
\end{align}\end{subequations}
These terms represent the interaction of the sources with the massive modes and contribute to the pre-exponential terms $\left( \partial_{h_+} S_{\rm src} \, \partial_{h_-} S_{\rm src} \right)$.
\begin{align}\label{eq.delhSsq}
	\partial_{h_+} S_{\rm src} \, \partial_{h_-} S_{\rm src} &= -\frac2{\gamma^2} \Bigg[ {\rm STr}\left[P_b\cdot W_{\mu_{k_1}}^{-+}\right] {\rm STr}\left[P_b W_{\mu_{k_2}}^{+-}\right] {\rm Tr}\left[ O X^{\mu_{k_1}}\right] {\rm Tr}\left[ O^\dagger X^{\mu_{k_2}} \right] \nonumber \\
	&  \qquad \quad + \left\{ \left( \frac E{2\gamma}  -i \sqrt{1-\frac {E^2}{4\gamma^2}} \right)^2 {\rm STr}\left[ P_b \cdot V^{++}_{\mu_{k_1}}\cdot W_{\mu_{k_2}}^{+-} \right] {\rm STr}\left[ P_b \cdot W^{-+}_{\mu_{k_3}}\cdot V_{\mu_{k_4}}^{++}  \right] \right. \nonumber \\
	&  \qquad \qquad \left.+ \left( \frac E{2\gamma}  + i \sqrt{1-\frac {E^2}{4\gamma^2}} \right)^2 {\rm STr}\left[ P_b \cdot W^{+-}_{\mu_{k_1}}\cdot V_{\mu_{k_2}}^{--} \right] {\rm STr}\left[ P_b \cdot V^{--}_{\mu_{k_3}}\cdot W_{\mu_{k_4}}^{-+} \right] \right\} \times\nonumber \\
	&\qquad\qquad\qquad\qquad {\rm Tr} \left[ O^\dagger X^{\mu_{k_1}} X^{\mu_{k_2}}\right] {\rm Tr} \left[ O X^{\mu_{k_3}} X^{\mu_{k_4}}\right] \Bigg]+ \cdots~,
\end{align}
where, ``$\cdots$'' represents terms that evaluate to zero.
An explicit computation shows that the terms proportional to ${\rm Tr} \left[ O^\dagger X^{\mu_{k_1}} X^{\mu_{k_2}}\right] {\rm Tr} \left[ O X^{\mu_{k_3}} X^{\mu_{k_4}}\right]$ evaluate to zero and the contribution of the massive modes to the correlation function is given by
\begin{equation}\label{eq.prop-mom}
	\frac1{2\pi^3\gamma^2\,\rho(E)} {\rm Re}\left[\frac1{ \epsilon(k) - i \omega}\right]  \times {\rm Tr}\left[ O X^{\mu_{k_1}}\right] {\rm Tr}\left[ O^\dagger X^{\mu_{k_1}} \right]~.
\end{equation}
where,
\begin{eqnarray}
\label{eq.massive-spectrum}
	\epsilon_k(E) &=& \frac 1 \pi  D \Delta(E)\left(S^{-1}(k)-1\right),\\[5pt]
	S(k) &=&\frac1{{N\choose 4}} \!\! \sum_{r=0}^{{\rm min}(k,4)}\!\! (-)^r {k\choose r} {{N-k}\choose{4-r}}~, \nonumber
\end{eqnarray}
is the spectrum of the massive modes in the SYK model at energy $E$.
Let us estimate this propagator for various massive modes. For all the above cases ($\mu\neq0$), assuming $|N/2-k|\gg1$, \footnote{$S(k)$ is an even function of $k$ around $N/2$.}
\begin{equation}
	S(k)^{-1} - 1 = \frac{8k}N + \mathcal O\left(\frac1{N^2}\right)~.
\end{equation}
This shows that all modes are strongly overdamped, with their masses being much larger than the many-body level spacing. Performing a Fourier transform over the frequency, $\omega$, to obtain the correlation function as a function of time,
\begin{equation}
	\frac1{2\pi^2\, \gamma^2 \, \rho(E)} \, \sum_{\{\mu_{k_1}\}} \exp\left[{- \epsilon_k(E) t}\right] \, {\rm Tr}\left[O X^{\mu_{k_1}}\right] {\rm Tr}\left[O^\dagger X^{\mu_{k_1}}\right]~.
\end{equation}
One can integrate over the spectrum to obtain the contribution of the massive modes to the correlation function,
\begin{align}
	\tilde C(t) &= \frac1{2\pi^2\, \gamma^2} \, \sum_{\{\mu_{k}\}} \int \frac{dE}{\rho(E)}\, \exp\left[{- \epsilon_k(E) t}\right] \, {\rm Tr}\left[ O X^{\mu_{k}}\right] {\rm Tr}\left[ O^\dagger X^{\mu_{k}}\right] \nonumber \\
	&= \frac1{D} \, \sum_{\{\mu_{k}\}} f\big(\epsilon_k(0) \, t\big) \, {\rm Tr}\left[ O X^{\mu_{k}}\right] {\rm Tr}\left[ O^\dagger X^{\mu_{k}}\right]~, \label{eq.Ct-intermediate}\\[5pt]
	&f(x) = 1 - \frac 2 \pi \int_0^{x} K_0(y)dy = 1- x \left[ K_0(x) L_{-1}(x) + K_1(x) L_{0}(x) \right]. \label{eq.Ct-intermediate2}
\end{align}
Above $K_n(x)$ is the modified Bessel function of the 2nd kind and $L_n(x)$ is the modified Struve function. The function $f(x)$ behaves as
\begin{equation}\label{eq.flimits}
	f(x) = \left\{
		\begin{array}{ll}
			1 - {\cal O} (x\ln x), & \qquad x \ll 1 \\
			\displaystyle \sqrt{\frac{2}{\pi x}}\,e^{-x}, & \qquad x \gg 1
		\end{array}
		\right.~,
\end{equation}
which yields a correct limit at $t \to 0$, i.e. 
\begin{equation}
\widetilde C(0) = \frac{1}{D} \sum_{\mu_k} {\rm Tr}\left[O X^{\mu_{k}}\right] {\rm Tr}\left[O^\dagger X^{\mu_{k}}\right] \equiv {\rm Tr}[ O O^\dagger].
\end{equation}
The last relation is a property of the `Fourier' transform in the Hilbert space which is based on the orthogonality relation
${\rm Tr} [X_\mu X_\nu^\dagger] = D \delta_{\mu\nu}$. Since,
$$
	\epsilon_k(0) \overset{ 1 \ll |N/2-k|}{\approx} \frac{8\gamma k}N = \frac{4k}{\sqrt N} J
$$
from \eqref{eq.flimits}, we know that the lightest modes correspond to $k=1,N-1$.

Contribution of the massive mode along with the ergodic answer, \eqref{eq.CRMT}, gives the full contribution to the correlation function,
$$
	C(t) = C_{\rm RMT}(t) + \tilde C(t)~.
$$
For the two-body hopping operators, $O = \gamma_i \gamma_j$, typical plots of the full correlation function $C(t)$ are shown in \autoref{fig.massive-massless}.

\subsection{Ergodic time}
Let's further estimate the ergodic time,  $t_{\rm er}$, at which the minimum of $C(t)$ is reached. Assuming that $O = X_\nu$ with $k=|\nu|$ and $|N/2-k|\gg1$ along with $\tilde t \ll 1$ the correlation function expressed via scaled time $\tilde t = t/t_H$ becomes,
\begin{equation}
	C(\tilde t) \approx D f\left(  \frac{16 D k}{N} \tilde t \right) + \frac {4\tilde t}{\pi}, \qquad \tilde t \ll 1.
\end{equation}
The minimum of this function is reached at time $\tilde t_{\rm er}$ satisfying
\begin{equation}
 \frac{16 D^2 k}{N} K_0\left(  \frac{16 D k}{N} \tilde t_{\rm er} \right) = \frac {4}{\pi}.
\end{equation}
Omitting further all constants of order unity and using the asymptotics of Bessel function at large arguments, the above equation is reduced to
\begin{equation}
\frac{D^2}{N}\frac{e^{-y}}{\sqrt{y}} \sim 1, \qquad y = \frac{16 D k}{N} \tilde t_{\rm er}.
\end{equation}
With $D=2^{N/2-1}$ this equation is solved by $y \simeq N \ln 2 - \tfrac 32 \ln N$, which leads to
\begin{equation}
\tilde t_{\rm er} = \frac{N^2 \ln 2}{16 D k} \qquad \Rightarrow \qquad t_{\rm er} \sim \frac {N^{3/2}}{Jk}.
\end{equation}

\subsection{Evaluation of energy integral}
Taking into account the energy dependence of $\rho(E)$ and $\epsilon_k(E)$, the energy integral which follows from \eqref{eq.Ct-intermediate} becomes
\begin{align}
	\widetilde{C}(t) &= \frac{1}{2\pi D \gamma} \sum_\mu \int\limits_{-2\gamma}^{2\gamma} \frac{dE}{\sqrt{1- E^2 /4\gamma^2}}  \exp\left(- \frac{\epsilon_k t}{\sqrt{1-{E^2}/{4\gamma^2}}} \right) |{\rm Tr}( O X^\mu )|^2 \nonumber \\
	&\overset{ E = 2 \gamma z}{=} \frac{1}{D} \sum_\mu f( \epsilon_k t) |{\rm Tr}( O X^\mu )|^2,
\end{align}
where the dimensionless function $f(x)$ reads
\begin{equation}
f(x) = \frac 1 \pi \int\limits_{-1}^{+1} \frac{d z}{\sqrt{1-z^2}} \exp\left(- \frac{x}{\sqrt{1-z^2}}\right) 
\overset{z= \tanh \theta}{=} \frac 2 \pi \int\limits_{0}^{+\infty} \frac{e^{- x \cosh \theta} }{\cosh\theta} \, d\theta.
\end{equation}
At the origin one finds $f(0)=1$. While for the derivative  one gets
\begin{equation}
f'(x) = - \frac 2 \pi \int\limits_{0}^{+\infty} e^{- x \cosh \theta}\, d\theta = - \frac 2 \pi K_0(x),
\end{equation}
from where the final result for the correlation function~\eqref{eq.Ct-intermediate2} directly follows.

\bibliographystyle{utphys}
\bibliography{refs}{}
\end{document}